\documentstyle[12pt,epsf,epsfig]{article}
\newlength{\extraspace}
\newlength{\extraspaces}
\newcommand{\be}{\begin{equation}}
\newcommand{\ee}{\end{equation}}
\newcommand{\bea}{\begin{eqnarray}}
\newcommand{\nn}{\nonumber}
\newcommand{\eea}{\end{eqnarray}}
\newcommand{\nd}[1]{/\hspace{-0.6em} #1}

\newcommand{\erf}{{\rm erf}}
\newcommand{\intR}{\int\limits_{-\infty}^\infty}

\def\lsim{\mathrel{\rlap {\raise.5ex\hbox{$ < $}}
{\lower.5ex\hbox{$\sim$}}}}

\newcommand{\pr}{\paragraph{}}

\def\gappeq{\mathrel{\rlap {\raise.5ex\hbox{$>$}}
{\lower.5ex\hbox{$\sim$}}}}

\def\lappeq{\mathrel{\rlap{\raise.5ex\hbox{$<$}}
{\lower.5ex\hbox{$\sim$}}}}

\begin{document}

\thispagestyle{empty}

\begin{flushright}
{\sc ACT}-14-96 \\
{\sc CERN-TH}/96--264\\
{\sc CTP-TAMU}-49-96 \\
{\sc OUTP}-96-57P\\
hep-th/9609238\\
September 1996
\end{flushright}
\vspace{.3cm}

\begin{center}
{\large\bf{ $D$-Brane Recoil Mislays Information}}\\[4mm]

{\bf John Ellis}\\ [2mm]
{\it Theoretical Physics Division, CERN, Geneva}\\[3mm]
{\bf N.E. Mavromatos}
\footnote{P.P.A.R.C. Advanced Fellow}\\[2mm]
{\it Theoretical Physics, 1 Keble Road\\
      Oxford, OX1 3NP, UK} \\[3mm]
{\bf D.V. Nanopoulos}\\[2mm]
{\it Center of 
Theoretical Physics, Department 
of Physics, Texas A \& M University,
College Station, TX 77843-4242, 
and HARC, The Mitchell Campus, Woodlands (Houston), TX 77381, USA}\\[15mm]

{\sc Abstract}

\end{center}

We discuss the scattering of a light closed-string state off
a $D$ brane, taking into account quantum recoil effects on the latter,
which are described by a pair of logarithmic operators. The light-particle
and $D$-brane subsystems may each be described by a world-sheet with an
external source due to the interaction between them. This perturbs each
subsystem away from criticality, which is compensated by dressing 
with a Liouville field whose zero mode we interpret as time. The
resulting evolution equations for the $D$ brane and the closed string
are of Fokker-Planck and modified quantum Liouville type, respectively.
The apparent
entropy of each subsystem increases as a result of the interaction
between them, which we interpret as the loss of information resulting
from non-observation of the other entangled subsystem. We speculate
on the possible implications of these results for the propagation of
closed strings through a dilute gas of virtual $D$ branes.

\noindent

\vfill
\newpage
\pagestyle{plain}
\setcounter{page}{1}
\stepcounter{subsection}
\section{Introduction.}

A new era in the study of string theory and black holes has
been opened up by Polchinski's realization~\cite{dbranes} 
that soliton backgrounds
in string theory can be described in a conformally-invariant way,
in terms of world sheets
with boundaries (thus incorporating 
open strings), on which Dirichlet boundary conditions 
for the collective target-space coordinates of the soliton are imposed.
One of the most fruitful applications of this $D$-brane
technology has been to black holes. In particular, many
authors~\cite{counting}  have demonstrated that the counting of 
quantum states of $D$ branes is equivalent to that of black hole
states~\cite{hawkbek}. Thus, it is now generally agreed that
black-hole entropy may be dissected into string states that are
in principle distinguishable, as we conjectured some time 
ago~\cite{emn,emnhair} on the basis of studies of two-dimensional stringy
black holes~\footnote{See also~\cite{KN}: a similar conjecture was made
later by
Susskind~\cite{susskind} on different, heuristic grounds.}. One of
the ways of distinguishing black-hole states is via generalized
Aharonov-Bohm measurements~\cite{emnmeas}, as has recently been
discussed~\cite{banks} in the context of $D$ branes.

\pr
We have shown recently~\cite{emnd} how $D$ branes emerge in a formulation
of the string analogue of the field-theoretical path integral, 
based on a treatment of non-critical string theory~\cite{emn}
in which the role of the target time variable $t$ is played by 
the Liouville field $\phi$~\cite{aben}, treated as a local
renormalization scale~\cite{emn,kogan}. We also pointed out 
in~\cite{emnd} several analogies between studies of $D$ branes
and two-dimensional string black holes~\cite{emn}. Our interest in 
the latter was largely motivated by the black hole information problem
and its possible implications for the effective quantum-mechanical
description of light-particle degrees of freedom~\cite{emn}. This problem has
also been studied extensively in the context of quantum gravity.
In particular, we have shown~\cite{emnw} that, at the one-loop level,
the entropy of a scalar field in
the presence of a four-dimensional black-hole background in
conventional general relativity diverges logarithmically with 
respect to the short-distance cutoff of the model, taken to be the
minimum distance from the black-hole horizon at which a `brick-wall'
boundary condition is imposed on the wave function of the scalar
field. We have interpreted these quantum divergences as reflecting
irreversible
temporal evolution associated with entropy production as the horizon
moves due to black-hole radiation.

\pr
The purpose of this paper is to initiate analogous studies in
the context of $D$ branes, with an examination of the quantum effects
of $D$-brane recoil during the scattering of a light closed-string
state, which requires a treatment of $D$-brane excitations.
The remarkably simple construction of Polchinski~\cite{dbranes}
opens the way to a $\sigma$-model description of 
such $D$-brane excitations, in which
the critical world-sheet string action 
is perturbed using appropriate 
boundary terms. We recall the form of the world-sheet
boundary operators describing
the excitation of a $D$ brane (see ~\cite{callan} and references
therein):
\be
   {\cal V}_D = \int _{\partial \Sigma}( y_i \partial _n X^i
+ u_i X^0 \partial_n X^i)
\label{dbraneop}
\ee
where $n$ denotes the normal derivative on the boundary of the world
sheet $\partial \Sigma$,
which has at tree level the topology of a disk of size $L$, and the
$X^i~,i=1,\dots $ denote the collective excitations of the $D$ brane,
which satisfy Dirichlet boundary conditions
on the world-sheet boundary:
\be
 X^i ({\rm boundary}) = 0,~i=1,\dots,
\label{fiveb}
\ee
whilst  $X^0$ is the target time variable which satisfies  
standard Neumann
boundary conditions: $\partial _n X^0 ({\rm boundary}) = 0$.
For simplicity, we will later be considering the case of 
a $0$ brane, with
the quantity $u_i $ in (\ref{dbraneop}) denoting its
velocity, and $y_i$ its initial position~\footnote{We do
not discuss here the more general case of curved
world volumes, where the simple Dirichlet boundary
conditions are known not to be conformally
invariant~\cite{dbranes,li}.}. In this case, the
operators (\ref{dbraneop}) describe shifts and
motion of the $0$ brane, and so can be
thought of as generating the action of the Poincar\'e group on the 
$0$ brane, with the $y_i$ parametrizing translations and the $u_i$
parametrizing boosts. In the general $D$-brane case, these
represent translations and boosts acting on the surface ${\cal S}$
of the $D$ brane.

\pr
First steps were taken in refs. \cite{recoil}
towards the quantum theory of the scattering 
of string states off a $D$-brane background.
This involves considering the motion of a closed string state
towards the (initially fixed) surface 
${\cal S}$, the latter viewed as a string 
soliton background, as shown in Fig.~1(a). In the general
$D$-brane case, the surface ${\cal S}$ divides
the target space-time into two regions. 
The closed string state 
is initially far outside the 
surface of the brane. At a certain moment,
say $X^0=0$, 
the incoming closed string state finds itself lying partly outside and
partly inside the $D$-brane surface. There are then two possibilities
to be considered: it may be either absorbed (Fig. 1(b)) or 
rescattered (Fig. 1(c)).
We seek below the appropriate
conformal field theory description 
of the latter case.

\pr
In general, quantum scattering
off the $D$ brane excites an open string state on the surface
of the $D$ brane, which in the scattering case of Fig.~1(c)
also emits another closed string state. The quantum excitation and
emission processes are both described by closed-to-open
string amplitudes, which are non-zero
in a world-sheet theory with boundaries. The open-string states
are excitations on the $D$-brane collective-coordinate surface. 
As was shown in 
ref. \cite{cardy}, such processes can be described
in terms of data of the bulk theory. As we show in more detail
below, tracing over such 
excitations results in a quantum modification of the 
Hawking-Bekenstein area law for the entropy, which
has been shown to hold in tree-level treatments of $D$ branes.
 
\pr
To discuss the evolution of entropy, it is
essential to treat correctly the quantum recoil of
the $D$ brane during the scattering process, which
we discuss in section 2. 
The treatment of recoil requires~\cite{recoil,kogmav,periwal,emnd,kogwheat}
 an operator with non-zero matrix
elements between different $D$- (in our case $0$-)brane states. 
This can be achieved in the impulse approximation 
by introducing a Heaviside function factor, $\Theta (X^0)$,
into the second operator in (\ref{dbraneop}), which
describes a $0$ brane that starts moving at
time  $X^0=0$. The initial position of the
$0$ brane at $X^0=0$
is assumed to be given by the $y_i$. 
To determine the precise form of 
the recoil operator in our case, we observe that the leading quantum
correction to the scattering of a closed string
off a $D$ brane is given by an annulus, as shown in Fig.~2(a). This
is divergent in the limit where the annulus is pinched, as shown in
Fig.~2(b)~\cite{recoil,kogmav,periwal,emnd}.
As was done previously for closed strings in the context
of two-dimensional black holes~\cite{emn}\footnote{We
recall that in the similar analysis~\cite{emn} 
of two-dimensional stringy black holes,
the r\^ole of recoil was played by the back-reaction operator
which described a change of state of the black-hole background.
In the underlying conformal field-theoretical 
description of ref. \cite{witt}, 
the corresponding operator
was the world-sheet instanton anti-instanton vertex~\cite{yung}.
The instanton vertex induces 
(target-space infrared) infinities in correlation functions
that are identical to those
arising at the torus (closed-string) level~\cite{emn}. 
In this context, the instanton coupling constant is proportional to the 
string coupling, and thus a one-string loop analysis 
corresponds to a
dilute instanton gas approximation, as adopted in~\cite{emn}.}, 
we seek a deformation of the $\sigma$-model 
that reproduces at the tree (disc) level these 
infinities arising from quantum string 
loop corrections, which are infrared in target space.
This determines the form of the quantum recoil,
which is described by a pair of logarithmic operators,
and one recovers momentum conservation as a 
consistency check of the approach\cite{emn}. 
In the $D$-brane case discussed here, the 
weakly-coupled string limit: $g_s \rightarrow 0$ 
corresponds correctly to a one-open-string-loop (annulus) analysis for a 
semi-classical (heavy) $0$ brane ($D$ particle), since the mass of the 
latter is $M_D \propto 1/g_s$ in natural string units. 

\pr
We argue in section 3 that this treatment of the recoil of the $D$ brane
using logarithmic operators~\cite{gurarie,tsvelik,kogmav}
entails a time evolution of the $D$-brane state that
mirrors what we found previously in the cases of
non-critical Liouville strings~\cite{emn} and of a scalar
field in the presence of a four-dimensional black hole~\cite{emnw}.
The interaction between the incident light-string state and the
$D$ brane is described by a source term in the world-sheet
$\sigma$ model which induces an
apparent departure from criticality. This is compensated by
non-trivial dynamics of the Liouville field, whose zero mode
is identified with target time $t$, i.e., the zero mode of $X^0$.
As we demonstrate explicitly in section 4 using the $D$-brane equation of
motion based on a world-sheet Wilson-Polchinski
exact Renormalization-Group equation~\cite{wilson}, 
the evolution of the $D$-brane state
is of diffusive Fokker-Planck type, which
necessarily entails entropy production. This loss of information
in the $D$-brane sector, i.e., transition to a mixed quantum-mechanical
state, is accompanied, because of quantum entanglement, by a
corresponding transition to a mixed state in the truncated external
light-particle system, as we discuss in section 5.
The time evolution of the scattering closed-string state 
does not have a simple Hamiltonian description,
but provides a new explicit realization of the modified quantum
Liouville equation of~\cite{ehns} within the general
non-critical string framework of~\cite{emn,kogan}. As we discuss in
section 6, if we speculate on the extension of these results 
to the propagation of a closed string state through 
``virtual $D$ brane foam'', 
the maximum possible
rate of entropy growth that we find is similar in magnitude
to that we estimated previously in the two-dimensional 
Liouville-string~\cite{emn}
and four-dimensional black-hole contexts~\cite{emnw}.

\section{Operator Treatment of $D$-Brane Recoil}
\pr
We now review briefly the treatment~\cite{periwal} of
the scattering of a closed string off a $D$ brane,
restricting ourselves for reasons of simplicity
to the case of point-like $0$ branes:
the extension to higher-dimensional $D$ branes is straightforward. 
For this purpose, we need to consider the annulus amplitude of
Fig.~2(a), where the crosses denote
closed-string vertex operators $V(k)$, which must be 
integrated over the propagating open string.
This computation may be performed using
the operator formalism, in which one evaluates
${\rm Tr}V(k_1)\Delta V(k_2)\Delta$, with $\Delta ^{-1} \equiv 
L_0 -1 $, where $L_0$ is the Virasoro operator~\footnote{Alternatively, 
one may work in the closed-string channel, in which case one introduces
a state $|B>$ into the closed-string Fock space, to impose
the appropriate boundary conditions on the end of the 
closed-string world sheet: we return later to this approach.}. The part 
of this computation that is relevant for our purposes 
is that due to the world-sheet zero modes~\cite{periwal}.
Writing $\Delta \equiv \int _0^1  dx x^{L_0-2}$, $L_0=2p^2 + N$,
where $N$ is the string level number, and picking out the $N=0$
part, we find the following contribution to 
the annulus amplitude:
\be 
{\cal A} =
\int dq <q|exp(-ik_{1}^0X^{0})x_1^{-2(p^0)^2}
exp(-ik_{2}^0X^{0})x_2^{-2(p^0)^2}|q>
\label{annulus}
\ee
where the superscript $0$ denotes target-time components, and 
$q$ is the modular parameter of the annulus. 
The trace over the zero modes yields the generic form~\cite{periwal} 
\be
{\cal A} \ni \delta (k_1^0 + k_2^0)\sqrt{\frac{1}{{\rm log}(x_1)}}
f(x_2, k_2^0) 
\label{result}
\ee
which conserves momentum in the light-state sector,
i.e., does not include any $D$-brane recoil momentum.
The $\delta (k_1^0 + k_2^0)$ function 
arises from integrating over the zero modes of 
$\sigma$-model fields $X^0(z,{\bar z})$ in a 
standard fashion. There is also a corresponding 
$\delta (k_1 + k_2)$ over the space components, 
not written explicitly above, which comes 
from integrating over the world-sheet 
zero modes of the $X^i$ fields. 
The amplitude (\ref{annulus}, \ref{result})
is pathological, in the sense that it is divergent as
$x_1 \rightarrow 0$~\cite{periwal},
and requires regularization. It is this regularization 
that induces recoil effects which 
modify the $\delta (k_1+k_2)$ term 
in order to ensure momentum conservation in the presence 
of recoil.

\pr
The pathological behaviour in the limit $x_1 \rightarrow 0$
corresponds to the pinched-annulus configuration shown in
Fig.~2(b):
\be 
{\cal A} \ni
g_{s}\int _{x\sim 0}\frac{dx_1}{x_1\sqrt{8\pi{\rm log}(x_1)}}
A_{disk}(k_1,k_2) 
\label{result2}
\ee
where $A_{disc}(k_1,k_2) = <V(k_1)V(k_2)V^iV^i>$ is
the tree-level disc amplitude, with $\{ V_i \}$ 
denoting a complete set of eigenstates of the $L-0$ Virasoro 
operator, which includes $D$-brane Goldstone 
zero modes leading to dominant divergent 
contributions~\cite{recoil,kogmav,periwal,emnd}. 
To cancel this one-loop infrared divergence, one must add 
the following tree-level closed-string operator counterterm
\be
 \delta {\cal A} =   \int d^2z \partial _\alpha (f(X^0)\partial ^\alpha X^i)
\label{oper}
\ee
which contributes on the boundary. Its
form is determined by general properties 
of soliton backgrounds in string theory~\cite{kogmav},
and the function $f(X^0)$ in (\ref{oper}) is determined by requiring 
that the above operator 
reproduce the infinities of the annulus amplitude (\ref{result2}).
We refer the reader to~\cite{periwal} for details. 
For our purposes,  we restrict our attention to
the final expression for  the `impulse' operator~\cite{periwal}:
\be
V_{imp} \equiv
\int d^2z\thinspace \partial _\alpha ([u_i X^0 ]\Theta (X^0)
     \partial _\alpha X^i) =
 \int d\tau\thinspace  u_i X^0 \Theta (X^0)
     \partial _n X^i~; \qquad i=1, \dots 9.
\label{wrongrecoil}
\ee
The step-function operator in (\ref{wrongrecoil})
needs to be defined properly, and we adopt the
integral representation~\cite{kogwheat}:
\be
  \Theta_{\epsilon} (X^0) = -i\intR \frac{dq}{q - i\epsilon} e^{iq X^0}\quad,
\qquad \epsilon \rightarrow 0^+
\label{theta}
\ee
where $\epsilon$ is an infrared regulator parameter\footnote{The
sign of $\epsilon$ is given a physical origin in the next section.}. 
In (\ref{wrongrecoil}), the coupling $u_i$ denotes a change in 
the velocity of the brane, which is determined by
imposing overall conformal invariance of the
annulus and disc amplitudes~\cite{recoil}. 
The cancellation of tree and annulus divergences requires 
\be 
u_i =8\sqrt{2}\pi g_s(k_1 + k_2)_i
\label{velocity}
\ee
which expresses momentum conservation.
This interpretation of (\ref{velocity}) is consistent
with the fact that 
the soliton mass is proportional to $1/g_s$,
confirming the interpretation of $u_i$ as the $D$-brane velocity. 

\pr
Analysis of the operator product 
of the operator (\ref{wrongrecoil}) reveals that it is a 
{\it logarithmic } operator~\cite{gurarie}
\be 
     V_{imp}(x)V_{imp}(0) \sim {\rm log} x/|x|^2 
\label{log}
\ee
and can be decomposed into one of a {\it pair} of logarithmic 
operators $C$ and $D$~\cite{gurarie,tsvelik}, namely $D$ in the notation 
of \cite{kogwheat}. 
To identify the pair
let us concentrate, 
for convenience, on the $X^0$-dependent parts.  
The $D_\epsilon$ recoil operator is identified as~\cite{kogwheat}
\be 
D_\epsilon \equiv X^0 \Theta (X^0) 
\label{dopera}
\ee
The $C_\epsilon $ operator 
 can be found by studying the
operator product of $D_\epsilon$ 
with the stress-energy tensor:
\be
T(w)D_{\epsilon}(z)= ~\frac{-\epsilon^2/2}{(w-z)^2} D_{\epsilon}
+ \frac{1}{(w-z)^2}\epsilon\Theta(X^0)
\label{opetd1}
\ee
This enables us to identify the $C_\epsilon $ operator, using
the general properties of such logarithmic pairs~\cite{gurarie,tsvelik}:
\be
C_{\epsilon} =
 \epsilon  \Theta_{\epsilon} (X^0) 
\label{cop}
\ee
Thus, the correct $D$-brane recoil operator is~\cite{kogwheat}:
\bea
{\cal V}_{rec} =
\int d\tau ~ [y_i C_{\epsilon}(X^0) \partial _nX^i + 
 u_iD_{\epsilon}(X^0)
\partial _n X^i ] 
\label{recoil}
\eea
The analysis of ref. \cite{kogwheat} showed
that the degenerate operators 
$C_\epsilon$ and $D_\epsilon$ have conformal dimension 
$\Delta=-{\epsilon^2\over 2}$, which is negative and 
vanishing in the limit $\epsilon \rightarrow 0^+$. 
This means that the impulse  operators $V_{rec}$ 
(\ref{recoil}) are {\it relevant}
in a renormalization-group sense for any non-zero $\epsilon$,
since, due to their $\partial _n X^i$ parts, their anomalous 
scaling  
dimension is $\Delta -1 +1 =-{\epsilon ^2\over 2}$.
This is intrinsic to the nature of logarithmic operators,
which appear on the border line between conformal
field theories and general renormalizable two-dimensional
field theories \cite{kogmav},\cite{emnd},\cite{gkf}. 
These relevant deformations in the recoil problem lead
to a {\it change} in the background $0$-brane
state~\cite{kogmav,emnd}, with the physical consequences
discussed in the next section.

\pr
Explicit expressions for the one- and two-point functions
of the operators  $D_{\epsilon}(X^0)$  and 
$C_{\epsilon}(X^0)$  appearing in (\ref{recoil}) were
derived in~\cite{kogwheat}. It is sufficient for
our purposes here to quote the results for the two-point functions:
\bea
&~&<C_\epsilon (z)C_\epsilon (0)> 
\sim -\epsilon^2\sqrt{\frac{\pi}{\alpha}} 
\intR \frac{dq}{(q^2+\epsilon^2)}
e^{-2\eta q^2\log|z/a|^2} \nn\\
&~&=-\epsilon^2  \pi\sqrt{\frac{\pi}{\epsilon^2\alpha}} 
e^{2\eta\epsilon ^2 \log|z/a|^2} \left(1-\erf\left(\epsilon 
\sqrt{2\eta \log|z/a|^2}\right)\right)\nn\\
&~&\stackrel{\epsilon\to 0}{\sim} 0+O(\epsilon^2)
\label{cc3}
\eea
\newcommand{\ggg}{2\eta\epsilon^2 \log|z/a|^2}
\bea 
&~&<C_\epsilon (z)D_\epsilon (0)> 
\sim -\frac{\epsilon}{2}\sqrt{\frac{\pi}{\alpha}}
\frac{\partial}{\partial \epsilon} \intR \frac{dq}{q^2 + \epsilon ^2}
e^{-2\eta q^2 \log|z/a|^2}\nn \\
&~& = {\pi\over2}\sqrt{{\pi\over\epsilon^2\alpha}}
\bigg\{e^{\ggg}\left(1-4\eta\epsilon^2 \log|z/a|^2\right)
\left(1-\erf\left(\sqrt{\ggg}\right)\right)\nn\\
&~&\,\qquad\qquad\qquad+2\sqrt{{\ggg\over\pi}}\,\bigg\}\nn\\
&~&\stackrel{\epsilon\to 0}{\sim}{\pi\over2}\sqrt{{\pi\over\epsilon^2\alpha}}
\left(1-\ggg\right)
\label{cd}
\eea
\bea 
&~&<D_\epsilon (z)D_\epsilon (0)> =
\frac{1}{\epsilon^2} <C_\epsilon (z)D_\epsilon (0)> \nn \\
&~&\stackrel{\epsilon\to 0}{\sim}{\pi\over2}\sqrt{{\pi\over\epsilon^2\alpha}}
\left({1\over\epsilon^2}-2\eta\log|z/a|^2\right)
\label{dd}
\eea
We have denoted by $\eta$ is the signature of the target metric:
\be
\hbox{Lim}_{z \rightarrow 0} < X^0(z) X^0(0) > \simeq \eta \log |a/L|^2
\label{xx}
\ee
where $a$ is a conventional ultraviolet regulator parameter on the
world sheet and $L$ is an infrared regulator parameter which
characterizes the size of the world-sheet annulus.
We see explicitly that in the limit 
\be
\epsilon \rightarrow 0, \qquad
  \epsilon ^2 \log|L/a|^2 \sim O(1)
\label{epsilonlog}
\ee
we obtain the canonical two-point  correlation functions of a pair
of logarithmic operators~\cite{gurarie}, 
with one exception - the singular $1/\epsilon^2$ term in 
$<DD>$~\footnote{Note that the singularity structure at small $\epsilon$
is the same in
connected correlation functions~\cite{kogwheat}.}. The relevance of this 
singular term will become apparent in the next section, when
we discuss the equation of motion of the semiclassical
$D$-brane wave function~\cite{wilson,rey,emnd}.

\section{Renormalization-Group Rescaling and Time}

\pr
We now discuss in more detail renormalization-group rescaling
on the world sheet, with the aim of establishing a connection
with the target time variable. The value of the numerical
constant in (\ref{epsilonlog}) is a free parameter 
which we may choose it at will, with
the differences between different choices being reabsorbed in the
redefinition of nonleading $\log z$ terms.
It was argued in ref.~\cite{kogwheat} that the
most natural choice is  (\ref{epsilonlog}), i.e. 
\be
{1\over\epsilon^2} = 2 \eta \log|L/a|^2
\label{relation}
\ee
which leads to the following singularity structure:
\bea   <C_{\epsilon}(z) C_{\epsilon}(0) > &\sim& 0 + O[\epsilon^2] \nn \\
   <C_{\epsilon}(z) D_{\epsilon}(0) > &\sim & 1 \nn \\
    <D_{\epsilon}(z)D_{\epsilon}(0)> &\sim& -2 \eta \log|z/L|^2
\label{CD}
\eea
up to an overall normalization factor.

\pr
We now consider a finite-size scale transformation
\be
L  \rightarrow L' = L e^{t}
\label{fsscaling}
\ee 
of the only type which makes
physical sense for the open string world-sheet. 
The relation (\ref{relation}) between
$\epsilon$ and $L$ entails the following transformation
of $\epsilon$~\footnote{Note that, if $\epsilon$ is infinitesimally small,
so also is $\epsilon'$ for any finite $t$.}:
\be
\epsilon^2  \rightarrow \epsilon'^2 =
 \frac{\epsilon^2}{1 + 4\eta \epsilon^2 t}
\label{epsilontransform}
\ee
We deduce from the scale dependences of the correlation functions
(\ref{CD}) that the corresponding transformations of
$ C_{\epsilon}$ and $D_{\epsilon}$ are:
\bea
D_{\epsilon} &\rightarrow& D_{\epsilon'} =
D_{\epsilon} - t C_{\epsilon} \nn \\
C_{\epsilon} &\rightarrow& C_{\epsilon'}= C_{\epsilon}
\label{correspond}
\eea
We emphasize that this transformation law is
{\it unambiguous}, being in particular {\it independent} of
the signature parameter $\eta$. 

\pr
The corresponding transformation laws for
the couplings $y_i$ and $u_i$, which are conjugate to 
$ D_{\epsilon}$ and $C_{\epsilon}$  in the recoil expression
(\ref{recoil}), are
\be
 u_i \rightarrow u_i~~,~~y_i \rightarrow y_i + u_i t 
\label{scale2}
\ee
This is consistent with the interpretations of $u_i$
as the velocity after the scattering process
and $y_i$ as the spatial collective coordinates 
of the brane, if and only if the scale-change parameter $t$ is
interpreted as the {\it target Minkowski time}\footnote{It 
is apparent from the scale transformation (\ref{correspond}) 
that
even if one started with the $D_\epsilon$ operator 
alone, a $C_\epsilon$ operator would have been induced,
as required if the pair (\ref{recoil}) is to
describe correctly $D$-brane recoil, as in (\ref{scale2}).},
and is therefore to be identified with the zero mode of $X^0$.
In this analysis
we have assumed that the velocity $u_i$ is small, as is
appropriate in the weak coupling r\'egime studied 
here (\ref{velocity}). The $D$-brane $\sigma$-model formalism
is completely relativistic, and we anticipate that a complete
treatment beyond the one-loop order discussed here will
incorporate correctly all relativistic effects, including
Lorentz factors wherever appropriate.

\pr
This observation that
a world-sheet scale transformation leads to the
target-time evolution of the $D$ brane, 
once recoil is taken into account, is
in the same spirit as was proposed previously in the
context of Liouville strings~\cite{emn,kogan}. There also
an identification was made between target time and a
world-sheet renormalization scale, provided in that case
by the zero mode of the Liouville field. 

\pr
We now provide technical support for a similar
identification in the present model 
of the zero modes of the target time
variable $X^0$ (\ref{dbraneop}) and the
Liouville scale field $\phi$~\footnote{In our interpretation~\cite{emn}, $\phi$ is a local
renormalization scale on the world sheet, and its zero mode
is identified with $t = \log |L/a|$.}, which is defined by
$\gamma_{\alpha \beta} = e^{\phi} \delta_{\alpha \beta}$ for the
disc topology, where $\gamma_{\alpha \beta}$ is the world-sheet
metric. We first rewrite the boundary recoil operators 
(\ref{recoil}) as world-sheet bulk operators 
\be 
   V_i = \int _\Sigma d^2z \partial _\alpha (y_i (X^0) \partial ^\alpha X^i )
\label{bulk}
\ee
Since they have non-vanishing
anomalous dimension $-\epsilon ^2/2$, the deformed theory 
is non-critical. Local scale (conformal) 
invariance may be restored in the usual
way by dressing (\ref{bulk}) with the Liouville field 
$\phi (z,{\bar z})$~\cite{DDK}:
\be 
  V_i^L \equiv 
\int _\Sigma d^2z e^{\alpha _i \phi (z,{\bar z})}
\partial _\alpha (y_i (X^0) \partial ^\alpha X^i )
\label{bulkliouv}
\ee
where $\alpha _i (\alpha _i + Q) =-2\delta _i $:
$Q$ is the central-charge deficit in the unperturbed theory and
$\delta _i=-\epsilon ^2/2$ the anomalous dimension. 
In our case, $Q=0$, consistent with fixed Dirichlet boundary 
conditions~\cite{li}, 
so that $\alpha _i =\epsilon$.     
Partial integration of (\ref{bulkliouv}) then leads to 
\be
V_i^L = \epsilon \int _{\Sigma} d^2z \int \partial _\alpha \phi
\partial^\alpha X^i 
y_i (X^0) + \int _{\partial \Sigma} d\tau e^{\epsilon \phi (\tau)}
y_i(X^0) \partial_n X^i
\label{expressionliouv}
\ee
We now use the fact (\ref{scale2})
that $y(X^0) = u_i X^0\Theta (X^0)$, and the 
integral representation (\ref{theta}), to observe that the
second term in (\ref{expressionliouv}) contains a factor
$e^{i(q X^0 - i \epsilon \phi)}$. Identifying
$\phi = X^0$, we may write
\be
\Theta _\epsilon (X^0) e^{\epsilon X^0} = 
{1 \over i}\int^{\infty}_{-\infty} \frac{dq}{q-i\epsilon}
e^{i(q-i\epsilon)~X^0} = {1 \over i} \int^{\infty - i \epsilon}_{- \infty
- i
\epsilon} \frac{d\omega}{\omega} e^{i\omega~X^0}
\label{thetaliouv}
\ee
which implies that, as far as the regulated $\Theta _\epsilon (X^0)$
is concerned, 
the effect of the Liouville dressing and the identification 
of $X^0$ with $\phi$ is equivalent to setting $\epsilon \rightarrow 0$
in (\ref{theta}), corresponding to the restoration of conformal 
invariance. Thus the promotion of the local renormalization scale $t$
to a Liouville field and its identification with the time $X^0$ are
consistent, if the following target-space metric $G_{MN}$ 
derived from (\ref{expressionliouv}) is considered:
\be
 G_{00}= 1 \qquad ; G_{ij} = \delta _{ij} \qquad 
     G_{0i} = G_{i0} = \epsilon y_i (X^0) = \epsilon u_i X^0 \Theta (X^0)   
\label{metrictarget}
\ee
where $0$ indices denote `time' components, and latin indices denote 
spatial components. 
In our discussion so far, we have 
ignored the effects of the $C-D$ mixing discussed previously
(\ref{correspond}). 
Indeed, due to this 
mixing, the effects of the $C_\epsilon$ recoil operator 
cannot be ignored~\cite{kogwheat}. 
Such effects contribute a term $\epsilon y_i \Theta (X^0)$ 
in the world line of the $0$ brane 
$y_i(X^0)$, with $y_i$ the initial location of the brane. 
According to the discussion in ref. \cite{kogwheat} and in the 
Appendix, 
such terms express quantum 
fluctuations in the location of the brane due to 
the excitation of {\it stringy} modes. 
Thus (\ref{metrictarget}) becomes
\be
      G_{0i}=G_{i0} =\epsilon (\epsilon y_i + u_i X^0)\Theta (X^0)
\label{metriccomplete}
\ee
The off-diagonal metric component $G_{0i}$ 
appears to have 
a  discontinuity at $X^0 = 0$. 
As we shall argue below, the $\epsilon y_i$-dependent   
parts of (\ref{metriccomplete})
lead to a curvature-singularity
at $X^0=0$ which will be $crucial$ for the information-loss problem.
\pr
The metric (\ref{metriccomplete})
is consistent with conformal invariance of the $\sigma$ model, 
as can be seen by considering the derivative with respect
to the (local scale) $X^0$: 
\be 
\frac{dG_{0i}}{d X^0}= \epsilon [ \partial _0y_i(X^0) ]= 
\epsilon [u_i \Theta (X^0)
+ (\epsilon y_i + u_i X^0) \delta (X^0)]
\label{scalecond}
 \ee
The right-hand side of (\ref{scalecond}) may be identified 
as a generalized coordinate transformation in target space
of the form $\nabla_{(0} \chi_{i)} (X^0)$:
$\chi_i \sim \epsilon (\epsilon y_i +  
u_i X^0) \Theta (X^0)$, where the parenthesized 
index is symmetrized. Equation
(\ref{scalecond}) expresses the condition for conformal 
(Weyl) invariance of the respective $\sigma$ model, 
in the usual way, and demonstrates the consistency 
of the identification of $X^0$ with the Liouville scale\footnote{A 
similar situation occurred in the ($1+1$)-dimensional 
black hole case~\cite{emn}.}. We note that (\ref{scalecond})
contains a singularity of the $\delta$-function type,
corresponding to the discontinuity in the metric (\ref{metrictarget}).
There is a corresponding singularity in the Riemann
curvature scalar at $X^0=0$:
\bea
&~&R \ni -\epsilon ^2 y^i (X^0) \partial _0 y^i (X^0) + \dots \ni 
-\epsilon ^2 
y^i (X^0) y^i (X^0) \delta (X^0) + \dots = \nn \\
&~&-\epsilon^2 (\epsilon y_i + u^i X_0)^2
\delta (X^0)
\Theta (X^0) + \dots
\label{scalrcurv}
\eea
where the temporal coordinate is $X^0$, and 
we can regard $\epsilon y_i + u_iX^0 \equiv \xi_i$ as a
Galilean-transformed spatial
coordinate, as discussed in more detail in section 5,
where we construct the $\sigma$ model that describes
the low-energy matter degrees of freedom.  
We stress once more that, 
as discussed in ref. \cite{kogwheat} and in the Appendix, 
the singular terms in (\ref{scalrcurv}) are linked to 
the $C_\epsilon $ (quantum) recoil operator~\cite{kogwheat}, 
and express quantum fluctuations  
in the location of the recoiling $D$-brane due to the excitation 
of the virtual {\it stringy} effects discussed in section 4.

\pr
After identifying $X^0$ with a local renormalization scale
(zero mode of the Liouville field), we find that $D$-brane recoil
{\it induces a back reaction on the space-time geometry 
which results in a temporal singularity in the curvature scalar}.
In our view, this lies at the core of the information-loss problem 
associated with virtual $D$-brane excitations induced by the recoil at
time $X^0$, as we discuss in more detail in the next sections.  

\pr
\section{Recoil and Information Loss}

\pr
We now discuss the implications of the above recoil analysis
for the time evolution of the $D$-brane state. 
We start from the formalism of the
Wilson approach to the renormalization group, as
refined by Polchinski and applied previously in the
a $\sigma$-model approach to $D$ branes~\cite{wilson,rey,emnd}. 
We denote by $\Psi [X^i]$ the 
partition function of a $D$ brane with collective coordinates
$X^i$, computed at the disc and annulus (tree and one-loop) level,
which plays the r\^ole of the semi-classical wavefunction
of the $D$ brane~\cite{rey,emnd}. This obeys a
renormalization-group equation, which expresses
its scaling properties with respect to a generic
short-distance cutoff $\Lambda$ on the world-sheet~\cite{rey}:
\be
\Lambda {\partial \over \partial \Lambda} \Psi [X] = \frac{1}{2\!}
\int\int _{\partial \Sigma} d\tau _1 d\tau _2 
\Lambda {\partial \over \partial \Lambda} G (\tau_1,\tau_2) 
\frac{\partial ^2}{\partial X^i (\tau _1) \partial X^j (\tau_2)}
\Psi [X] + ...  
\label{wilson}
\ee
where $\partial \Sigma$ denotes the world-sheet boundary
and the two-point Green function 
$G(\tau _1, \tau_2)=<X^i(\tau _1)~X^j~(\tau _2)>$. 
The dots denote terms of the form 
$(\partial \Psi [X]/\partial X)^2$ , which represent 
interactions among $D$ branes, and are ignored to the 
order of accuracy of the present work. 
In our case, we identify $\Lambda$ with $L/a$
and $t = \log |L/a|$ (\ref{fsscaling}), so that
$\Lambda \partial/\partial \Lambda = \partial/\partial t$ and
the renormalization-group equation~\cite{rey} may be recast in
the form 
\be
{\partial \over \partial t} \Psi [X] = \frac{1}{2\!}
\int\int _{\partial \Sigma} d\tau _1 d\tau _2 
{\partial \over \partial t} G (\tau_1,\tau_2) 
\frac{\partial ^2}{\partial X^i (\tau _1) \partial X^j (\tau_2)}
\Psi [X] + ...  
\label{tform}
\ee
which has the form of a diffusion equation. In order
to evaluate the rate of change of the
entropy, we interpret $|\Psi [X]|^2$ in the natural way
as the probability density distribution for the $D$ brane 
to find itself in the position $X^i(t)$ at time $t$. 
Using this interpretation, (\ref{tform}) may be used to
derive a Fokker-Planck 
equation for the probability distribution 
${\cal P} \equiv |\Psi [X]|^2$:
\be 
\partial _t {\cal P} = D~\nabla _{X^i}^2 {\cal P} 
+ \dots  
\label{FP}
\ee
as we discuss in detail below. In~(\ref{FP}),
$D$ denotes a diffusion coefficient, the dots denote terms
representing interactions among the $D$ branes, which we will ignore, 
apart from a few speculative comments in the last section,
and the $X^i$ denote the (world-sheet-independent) zero modes
of the collective coordinates of the brane.
 
\pr
In order to derive~(\ref{FP}) and demonstrate that it is non-trivial,
we show that both of the factors in the integrand of~(\ref{tform})
are non-zero, using the analysis of the pair $C_{\epsilon},
D_{\epsilon}$ discussed in the previous section. We start with
$G(\tau_1, \tau_2)$, which depends on $t$ only via its
divergent part, which is due to the (world-sheet-independent) zero modes 
on the annulus. These divergent parts can readily be
computed in the $\sigma$-model approach by noting that 
the collective coordinates of the $D$ brane 
depend on the cut-off scale,
as a result of the shift (\ref{scale2}) due to 
$C$-$D$ operator mixing (\ref{correspond}). 
Thus the world-sheet-coordinate-independent 
logarithmic divergences in $G(\tau _1, \tau_2)$ 
can be studied~\cite{emnd} by replacing 
$X^i (\tau _1)$ by the boosted coordinate 
$u_i X^0(\tau _1) \Theta (X^0)$, i.e. the $D_\epsilon$ recoil 
operator itself. The calculation then reduces to the
evaluation of the free two-point function of this operator,
which was given in ref.~\cite{kogwheat}, and 
quoted above (\ref{dd}):
\be 
G(\tau _1, \tau _2) |_{divergence} \sim u_i^2
<D_{\epsilon } (x) D_{\epsilon} (x') >_{divergence} = u_i^2
log|L/a|^2 
\label{diver}
\ee
This leads to a parametrization of $G(\tau _1, \tau _2)$ 
by an effective diffusion coefficient $D$:
\be 
G(\tau_1, \tau_2) \simeq 2D t: \, t \simeq log|L/a|
\label{defd}
\ee
with 
\be
D =  u_i u^i
\label{dformula}
\ee
which we substitute into~(\ref{tform}) to obtain
\be
\partial _t \Psi [X] = u_i^2
<\int \int _{\partial \Sigma} 
d\tau _1 d\tau _2 {\partial^2 \over \partial X^i(\tau_1) \partial
X^j(\tau_2)} \Psi[X] + \dots >
\label{intermediate}
\ee
We must now verify that the integrand in (\ref{intermediate}) 
is also non-vanishing.

\pr
To see this, we first use the form (\ref{recoil}) of the $D$-brane
recoil operator to re-express this integrand as
\be
{\cal A}_2 \equiv <\partial _n X^i (\tau _1) \partial _{n'} 
X^j (\tau _2) \delta _{ij}>_{annulus}
\label{pole}
\ee
in the absence of any closed-string tachyon perturbation,
where the expectation value is taken in the unperturbed
$D$-brane $\sigma$ model, without $C,D$ deformations. 
The two-point function on the annulus in (\ref{pole}) is that
of the Goldstone-mode operators $\int_{\partial \Sigma} d\tau 
\partial _nX^i (\tau) $ and is proportional to the
shift in the mass of the Goldstone 
mode due to loop corrections~\cite{recoil}. However, since the Goldstone 
mode expresses the spontaneous breaking of an exact global symmetry 
in target space, namely translation invariance~\cite{kogmav,emnd}, 
its mass should be strictly 
zero. Indeed, the contour deformation technique applied 
in~\cite{recoil} confirms this, since the only 
contribution to this two-point function comes from 
the tachyonic pathologies of the bosonic string, which are
absent in supersymmetric theories. 
Thus, in the absence of matter deformations, the Wilson 
equation (\ref{tform}) has only trivial content.

\pr
However, equations (\ref{tform}) and (\ref{FP}) are
non-trivial when matter deformations are included.
To see this, 
we employ the effective tree-level approach
to recoil, where one deforms  the $\sigma$ model 
by the world-sheet boundary term (\ref{recoil}),
and performs a perturbative expansion in powers 
of the tachyon deformations. Specifically, at first order we
replace (\ref{pole}) by the three-point amplitude 
\be
{\cal A}_3 \equiv <\int _{\partial \Sigma} d\tau _1 \int _{\partial
\Sigma}
d\tau _2 \partial _n X^i (\tau _1) \partial _{n'}X^i (\tau _2) 
\int _{\Sigma} d^2\sigma  T(k) e^{ik_M X^M (\sigma) }>
\label{3point}
\ee
evaluated using a free $\sigma$-model action
on a world sheet with boundaries, which is the leading-order 
contribution of matter deformations in the weak-tachyon approximation
to the right-hand-side of (\ref{tform}). 
The non-zero contributions that
lead to diffusion arise when the bulk tachyon deformation 
approaches the boundary of the world sheet. 
In such a case, we expect the following boundary operator product 
expansion~\cite{dieh,cardy} to hold:
\be 
\Phi (z, {\bar z}, y) 
\sim \sum _{i} (2y)^{\Delta _i - x_\Phi}C^a _{\Phi\psi _i}
\psi _i (y) 
\label{boundary}
\ee
provided that the set of boundary conditions $a$ does not break
conformal symmetry. In (\ref{boundary}), $z,{\bar z}$ denote 
world-sheet (bulk) coordinates, 
the $y$ are world-sheet boundary coordinates, 
$\Phi (z,{\bar z})=e^{ik_MX^M(z,{\bar z})}$
is the bulk deformation by the closed-string tachyon, 
$x_\phi$ is its scaling dimension, and 
$\{ \psi _i (y) \}$ is a (complete) set of boundary operators
with scaling dimensions $\Delta _i$. 
In theories where logarithmic operators are absent, it was shown 
in~\cite{cardy} that the open-to-closed-string 
O.P.E. coefficients $C^a_{\Phi\psi} $ could be
re-expressed in terms of bulk data, in particular 
the coefficients $C^i_{jk}$ that arise in the 
ordinary bulk O.P.E.. An extension of this analysis to
general logarithmic operators is an important technical 
issue that falls beyond the scope of the present work~\cite{kogwheat2},
but we are able to establish that there is a 
non-vanishing coefficient in our case.

\pr
The boundary operator $\psi$ of interest to us is the $D$ operator
discussed in the previous sections. To see that the $C^a_{\Phi D}$ 
O.P.E. coefficient is indeed non-zero, we 
consider the O.P.E. between two bulk operators 
\be 
   \Phi (z,{\bar z})\Phi (0,0) \sim C_{\Phi\Phi}^I V_I 
\label{ope}
\ee
where $\{ V_I \}$ denotes a complete set of bulk (closed-string) 
operators. In the $D$-brane case, these include the
bulk version of the $D$-brane recoil operator (\ref{wrongrecoil}). 
The one-point function on the disc of this operator 
is divergent near the boundary of the disk~\cite{periwal}~\footnote{This
is cancelled by divergences in
the string amplitude on the annulus, as discussed at the 
beginning of section 2.}. Parametrizing the world-sheet
distance from the boundary by $i \zeta / 2$, we have
\be 
    <X^0 (i \zeta / 2) \Theta (X^0(i\zeta /2))> \sim \int \frac{dq}{2\pi}
\frac{1}{q^2} (\zeta)^{-q^2} \sim \sqrt{-{\rm log}\zeta}
\label{divoper}  
\ee
Since $\zeta$ is a short-distance scale on the world sheet,
we identify $-\log \zeta \sim \log |L/a|$, and then
the dominant term in the
$D$-brane two-point function of (\ref{ope}) is
\be 
   <\Phi (z,{\bar z})\Phi (0,0)>_{z \sim 0} \sim C_{\Phi\Phi}^{D} 
\sqrt{{\rm log}|L/a|} 
\label{divbulk}
\ee
Consider now the case where both 
$\Phi $'s in (\ref{ope}) lie very close to  the boundary of the disc,
which for convenience we parametrize by the real axis of the upper half 
of the complex plane. In this case, for each of the $\Phi $ one has the 
boundary O.P.E. expansion (\ref{boundary}). The 
separation of the two operators projected on the boundary will also 
be small, so one may use the O.P.E. on the boundary for the 
boundary excitations. We see 
from (\ref{dd}) that the dominant (divergent) contributions 
in that expansion come from the boundary $D$ operator, whose two-point 
function is divergent as ${\rm log}|L/a|$:
\be 
<\Phi (z,{\bar z})\Phi (0,0)>_{z \sim 0} \sim (C_{\Phi D}^{a})^2
{\rm log}|L/a|
\label{boundiv}
\ee
Equating the divergent contributions of (\ref{boundiv})
and (\ref{divbulk}), one obtains 
the relation 
\be 
(C^a_{\Phi D})^2 \sim \sqrt{\frac{1}{{\rm log}|L/a|}}~C^D_{\Phi\Phi}
\label{relation2}
\ee
The O.P.E. coefficient $C_{\Phi\Phi}^D$ on the right-hand side 
of (\ref{relation2}) encodes the amplitude for an in-state 
tachyon to scatter off the $D$-brane into an out-state tachyon, 
including recoil, which is known to be
non-zero~\cite{recoil,kogmav,kogwheat,emnd}. 
Equation (\ref{relation2}) implies that there is a
non-trivial `trapping' amplitude, $C_{\Phi D}^a$, 
which therefore contributes to the diffusion coefficient in
(\ref{FP}), and hence to the information loss via quantum recoil
operators. The above calculation  is incomplete, in that
it was based on an analysis of the leading divergences of the
recoil operators of the $D$ brane~\cite{kogwheat},
and it remains to calcuate the proportionality coefficients
in (\ref{relation2}), 
as well as the subleading-divergence contributions of the open-to-closed
O.P.E. in a theory with logarithmic operators. This programme
falls beyond the scope of the present article~\cite{kogwheat2},
but is not needed for our purpose: we have established that
both terms in the integrand of (\ref{tform}) are non-vanishing,
and hence that the Fokker-Planck equation (\ref{FP}) is non-trivial.

\pr
The diffusion interpretation of (\ref{tform}, \ref{FP}) is
contingent on the absence of a factor $i$ between the two
sides of the equations, and one might wonder whether these
equations are correct as written, or whether they should be
Wick rotated. We would like to emphasize that the translation
laws (\ref{correspond}, \ref{scale2}) were derived
independently of the assumed signature $\eta$, and hence
that there is no justification for any Wick rotation of the time
variable in (\ref{tform}, \ref{FP})~\footnote{Indeed, 
when $\eta =-1$,
in which case $X^0$ has Minkowskian signature, there is a factor of $i$ 
on the right-hand-side of (\ref{pole}), as a result of the 
purely imaginary normalization factor $\sqrt{1/\epsilon ^2 \eta Log|L/a|^2}$
of the two-point function of the operator $D$ (\ref{dd}).
Such factors are, therefore, crucial.  
Their presence implies that, in the case $\eta =-1$, 
the target-space physical time should be identified with $it$, where
$t$ denotes the change (\ref{fsscaling}) in the world-sheet scale. 
This demonstrates that, with respect to the physical time, the equation 
(\ref{pole}) retains its diffusion form, whichever the 
signature $\eta$.}. 
The appearance of a diffusion equation can be traced back
to the presence in the recoil 
operators of a factor $\Theta (X^0)$~\cite{periwal,kogwheat}. 
In general, one should
also take into account the possibility that the $D$ brane
is moving initially, in which case one should use
the boundary operators (\ref{dbraneop}) {\it as well}.
In this case, there are trivial logarithmic operators
in the two-point function $G(\tau _1, \tau _2)$ of (\ref{wilson}),
coming from the 
free fields $X^0$ in the expression for the collective coordinates
$y(X^0)_i = y_i + v_i X^0$ of a $D$ brane moving with initial velocity
$v_i$. In the presence of both sets of operators,
the right-hand side of (\ref{wilson}) 
becomes {\it complex} for Minkowskian-signature $X^0$, as a result 
of the imaginary normalization factors appearing in the 
two-point functions of the $\Theta (X^0)$ operators, mentioned
above. This leads to an equation of motion which contains
a standard Hamiltonian term \`a la Schr\"odinger, corresponding to 
a freely-moving brane with finite velocity $v^i$, as well as
a diffusion term corresponding to the virtual 
excitations that appear after the time $X^0=0$.
We note in passing that 
the recoil treatment of \cite{recoil} used only 
eternally-moving $D$ branes, in contrast to the treatments 
in \cite{periwal,kogwheat}, in which
the membrane receives an impulse at the time $X^0$.
In this article, our interest is focused
on the diffusion aspect of the recoil, and we shall not
discuss further the Hamiltonian terms in the wave equation
of the moving brane. 

\pr
In view of the important implications of
(\ref{FP}), we now provide an alternative derivation that does not
use the above identification 
of the $\sigma$-model partition function with the 
semi-classical wavefunction of the $D$ brane.
This second derivation is based on the equivalence of the $D$-brane 
$\sigma$ model with a background gauge-field open string.
In this formulation, the boosted collective coordinates     
$X^i$ are viewed as `couplings' of a non-critical 
$\sigma$-model theory. The inclusion of higher genera
necessitates a summation over pinched handles
(cf, the prototypes in Fig.~2(b)), which entails
a natural quantization of these couplings~\cite{emnd},    
in close analogy to the wormhole calculus familiar from
higher-dimensional quantum gravity. As discussed earlier,
the spontaneous breaking of translational
invariance by the collective coordinates of the $D$ brane, 
which is a general feature of any string theory 
involving solitonic structures~\cite{kogmav},
is accompanied by the appearance of Goldstone zero modes.
As discussed previously in the literature~\cite{recoil,kogmav,emnd},
the propagation of these zero modes along thin tubes 
yields logarithmic infinities in loop amplitudes, 
which manifest themselves in bilocal structures on the world
sheet:
\be 
 {\cal B}_{ii} =\int d^2z V_i (z) \int d^2w V_i (w)
 \frac{1}{L_0+{\overline L}_0 -2}
\label{bilocal}
\ee
where the last factor represents the string propagator $\Delta _S$
on a degenerate handle, with the symbols $L_0, {\overline L}_0$
denoting Virasoro generators as usual. Inserting a
complete set ${\cal E}_{\alpha}$ of intermediate string
states, we can rewrite (\ref{bilocal}) as an integral
over the parameter $q \equiv e^{2 \pi i \tau}$, where
$\tau$ is the complex modular parameter characterizing
the world-sheet tube.
The string propagator over the world-sheet tube then reads
\be
\Delta _S \,=\,   \sum _\alpha   \int dq d{\overline q}
  \frac{1}{q^{1-h_\alpha} {\overline q}^{1-{\overline h}_\alpha}}
\{{\cal E}_\alpha (z_1)
\otimes (ghosts) \otimes
{\cal E}_\alpha (z_2) \}_{\Sigma_1 \oplus \Sigma_2}
\label{props}
\ee
where $h_\alpha, {\overline h}_\alpha $ are the
conformal dimensions of the states ${\cal E}_\alpha$.
The sum in (\ref{props}) is over all states
propagating along the long, thin tube connecting
$\Sigma_1$ and $\Sigma_2$, which are both equal to the
sphere in the simplest case of a degenerate torus.
As indicated in (\ref{props}), the sum over states must
include the ghosts, whose central charge cancels that of
the world-sheet matter theory in any critical string model.
States with $h_\alpha = {\overline h}_\alpha = 0$ may
cause extra logarithmic divergences in (\ref{props})
which are not included in the familiar $\beta$-function
analysis on $\Sigma$~\cite{recoil}.
This is because such states make contributions to
the integral of the form
$\int dq d{\overline q} / q {\overline q}$
in the limit $q \rightarrow 0$, which represents a
long, thin tube. We assume that such states are discrete in
the space of states, i.e., they are separated from other
states by a gap. In this case, there are factorizable
logarithmic divergences in (\ref{props}) which depend
on the background surfaces $\Sigma_{1,2}$, e.g., the
sphere in the case of the degenerating torus.
 
\pr
The bilocal term (\ref{bilocal}) can be cast in the form of
a local contribution to the world-sheet action, if one
employs the trick, familiar from the wormhole calculus, of
rewriting it as a Gaussian integral~\cite{wormholes,recoil}:
\be
  e^{{\cal B}_{ii}} \propto \int d\alpha
  e^{-\alpha _i^2 + \alpha _i\int V_i }
\label{wlocal}
\ee
where the $\alpha ^i$ are to be viewed as
quantum coupling constants/fields of the world-sheet
$\sigma$ model. In fact, the factor
$e^{-(\alpha _i)^2} $ must be replaced~\cite{wormholes,schmid2,emnd}
by a more general Gaussian distribution of width $\Gamma$:
\be
F(\alpha _i) =
 \frac{1}{\sqrt{2\pi}\Gamma} e^{-\frac{1}{2\Gamma ^2}(\alpha _i)^2}
\label{gauss}
\ee
The extra logarithmic divergences associated with degenerate
handles that we mentioned above, which are $\propto \hbox{ln} \delta$
where $q \sim \delta \sim 0$, have the effect of causing the
width parameter $\Gamma$ also to depend logarithmically on
the cutoff scale $\delta$:
\be 
   \Gamma \sim \hbox{ln}\delta 
\label{widthdiv}
\ee
Using the Fischler-Susskind mechanism~\cite{fiscsussk}
for cancelling string loop divergences, one should 
associate the divergence $\propto\hbox{log}\delta$ 
with a world-sheet cut-off scale $\hbox{log}\epsilon$. 
Upon identification of the
cut-off scale with the Liouville field, and of the
latter with the target-time variable~\cite{emn,kogan}, we
infer that the distributions of the couplings $\alpha_i$ become
time-dependent~\cite{emn}.

\pr
In the particular case of $D$ branes, the couplings
which fluctuate are the boosted collective coordinates of the 
$D$ brane, including the recoil velocities
(or momenta). Redefining the `quantum' coordinates by
$X^i \equiv y^i + \alpha _i \sqrt{\hbox{log}\delta}$,
the above analysis shows that the quantized collective 
coordinates have a Gaussian `white-noise' distribution 
of the generic form: 
\be
F(\delta X^i_q)=[\frac{1}{2\pi <\delta X^I_q>^2}]^{1/2}
{\rm exp}(-\frac{(\delta X^(_q))^2}{2<(\delta X_q^i)^2>})
\label{gaussian}
\ee
where the $\delta X_q^i$ denote quantum fluctuations of the 
time-dependent collective coordinates. Due to the above-mentioned
divergent structure associated with zero modes, the 
width of this distribution has the form: 
\be 
<(\delta X_q^i)^2> \sim~D~\delta t \qquad 
t =D^{-1}\hbox{log}\delta
\label{gaussianshift}
\ee
where $\delta t$ denotes an infinitesimal fluctuation in target time, and 
the relative normalization $\hbox{log}\delta = D~t$
of the diffusion coefficient 
is dictated by analogy with the Liouville theory~\cite{aben,emn}.
This structure is very similar to that of an
inflationary scenario for string cosmology~\cite{emninfl}, which
is understood in the context of Liouville strings~\cite{emn}, to which
this theory is equivalent.

\pr
We are now in a position complete our 
second derivation of a Fokker-Planck 
equation for the  probability distribution ${\cal P}(X^i,t)$
in the `space' $X^i$~\cite{infl,emninfl}, using
the renormalization-group equation for the coupling $X^i(t)$
\be
                   \partial _t X^i(t) = u^i 
\label{string}
\ee
which we interpret as an equation of motion for the classical coupling
$X^i_c(t)$. We then decompose the value $X_c(t + \delta t)^i $
at a later time $t$ as
\be 
X_c(t + \delta t)^i = X_c(t)^i   + \partial _t X^i(t)\delta t 
+ \delta X_q^i (t) 
\label{quantum}
\ee
where the $\delta X_q^i$ denote the quantum fluctuations
that satisfy the white noise distribution (\ref{gaussianshift}).
The probability distribution 
${\cal P}$ may then be written as~\cite{infl,emninfl}:
\be
{\cal P}(X^i,t) = \int dX_c^i F(\delta X_q^i)P(X_c,t) 
\label{prob}
\ee
from which (\ref{FP}) is derived in a straightforward manner,
as discussed in \cite{emninfl}~\footnote{We note in passing
that {\it if} the Fokker-Planck equation holds also in the
presence of $D$-brane interactions, {\it then} target-space 
diffeomorphism invariance restricts~\cite{goldin} the form of the
equation for the temporal evolution of the $D$-brane wave function
to a non-linear Schr\"odinger equation:
\be
i \hbar \partial _t \Psi = {\cal H} \Psi + i \hbar D \nabla^2_i \Psi
+ i D \hbar | \nabla_i {\rm log} \Psi |^2 \Psi
\ee
This equation might be useful for the quantization of 
interacting $D$-branes.}. This provides
an independent {\it a 
posteriori} justification for identifying the $D$-brane
$\sigma$-model partition function with 
the semi-classical wavefunction in target space. 

\pr
Having derived a Fokker-Planck equation for the probability 
distribution, we now consider the rate of change of the entropy,
defined as:
\be 
S=-k_B \int [DX_i] {\cal P}(X,t)\hbox{log}{\cal P}(X,t)
\label{entr}
\ee
where $[DX]$ denotes an appropriate functional integral
in a $\sigma$-model sense.
It is straightforward using (\ref{wilson}) to derive~\cite{emninfl}
\be
\partial _t S = \int [DX] u_ju^j \frac{|\nabla _{X^i}{\cal P} |^2}{{\cal P}}
+ \dots 
\label{rate}
\ee
where we work in a system of units such that $k_B=1$, and 
the $\dots $ indicate terms due to interacting $D$ branes,
that have not been discussed here.
The rate of change (\ref{rate}) is not only non-vanishing, but 
obviously monotonically increasing.

\pr
The rate of evolution (\ref{rate}) of the $D$-brane entropy
may be estimated for asymptotically-large time $t$, using
the analysis of the logarithmic conformal 
field theory describing the recoil process that we developed
in sections 2 and 3. In a semi-classical treatment,
we may represent ${\cal P} = |\Psi |^2$, where the 
$\sigma$-model partition function $\Psi =e^{-F}$,
with $F$ denoting the effective $D$-brane action:
\be
\partial _t S = \int [DX] u_iu^i \frac{1}{{\cal P}}
| \nabla _{X^i} {\cal P} |^2 = \int[DX] |\Psi |^2 (u_iu^i) 
|\nabla _{X^i} F|^2
\label{intermsoff}
\ee
Then, viewing the collective coordinates $X^i$ as couplings of 
the $D$-brane $\sigma$ model, using the $t$ dependence of $X^i$
expressed in (\ref{string}), and noting that no other couplings
of the $D$-brane $\sigma$ model depend on $t$ to this order,
we may write
\be
\partial _t S = \int [DX] |\Psi [X]|^2 |\partial _t F|^2
\label{rewrite}
\ee
Using now the renormalizability of the effective action $F$, 
and the Zamolodchikov $C$-theorem~\cite{zam} which tells us that
\be
\partial _t F = - \beta^{y^i} G_{CC} \beta^{y^i}
\label{zamthis}
\ee
in this case, since (\ref{scale2}) $\beta^{y^i} = u^i$, $\beta ^{u_i} =0$,  
and recalling  
that (\ref{CD})
$G_{CC} \propto \epsilon ^2 \sim 1/t$, we rewrite (\ref{rewrite}) as
\be
\partial _t S \propto (u_iu^i)^2 \frac{1}{t^2} 
\label{zamol}
\ee
where we have normalized the wavefunction $\Psi[X^i]$ 
in the space of the collective coordinates.
The result (\ref{zamol}), which is valid for 
asymptotically-large times $t \rightarrow \infty$, 
indicates that the change $\Delta S$ in $D$-brane entropy produced over 
the entire duration of the scattering process $ \Delta \tau 
\lsim 1/\epsilon$ 
by the quantum fluctuations of the recoiling $D$ brane 
is {\it finite}. This is consistent with the 
$\delta$-function-type temporal singularity of
the effective space-time curvature scalar (\ref{scalrcurv}),
and the finite range of effective deviations from conventional
General Relativity in string theory.

\section{Evolution Equation for the Effective Theory of Light States}

\pr
We have shown in the previous section that the ``kick"
applied to the $D$ brane by the incident light closed-string
state provides it with a quantum-mechanical excitation that
leaves it with finite entropy $\Delta S$. Since the initial
combined (light particle, $D$ brane) state had no entropy,
we expect the final light-particle state also to have
non-zero entanglement entropy $\Delta S$. In string world-sheet
language, the interaction between the light state and the $D$
brane implies that their effective $\sigma$ models receive
equal and opposite perturbations. These act as
sources in the equations of motion of both subsystems,
making them both appear non-critical. Thus one must introduce a
renormalization scale also for the light-particle subsystem,
which must also acquire non-zero entropy, according to
general arguments~\cite{emn}.

\pr
To see this explicitly, we first review the
general formalism of ref. \cite{emn} for the description of closed
strings perturbed away from criticality by quantum-gravitational
backgrounds. Criticality is restored by the inclusion of
Liouville scaling factors and the target time is 
re-interpreted as the
zero mode of the Liouville field $\phi$,  which may then be identified
with that of the $D$-brane target-time coordinate $X^0$, as we have
shown also in the $D$-brane context in section 3. It follows from
the general analysis 
of~\cite{emn} that the effective time-evolution equation for the
light closed-string degrees of freedom takes the form
\be
\partial _t \rho = i [\rho, H] + i\nd{\delta H}\rho 
\label{ehns}
\ee
where $H$ is the light-state Hamiltonian, and the non-Hamiltonian
term $\nd{\delta H} \rho$ takes the generic form
\be
\nd{\delta H}\rho = \beta^i G_{ij} [g^j, \rho]
\label{generalemn}
\ee
Here the non-zero renormalization coefficients $\beta^i$
parametrize the apparent departure from criticality induced
by the couplings $g^j$, which become quantum variables when
higher-genus effects are taken into account, and 
$G_{ij} \propto <V_i V_j>$ is a suitable positive-definite
Zamolodchikov metric in the space of the
vertex operators $V_{i,j}$ corresponding to these couplings.
The presence of the non-Hamiltonian term~(\ref{generalemn})
then induces entropy production in the effective light-particle
theory:
\be
{dS^{light} \over dt} =  -\beta^i G_{ij} \beta ^j
\label{genentropy}
\ee
causing $S^{light}$ to vary monotonically if any $\beta^i \ne 0$
and $G_{ij}$ has definite sign~\cite{zam}.
As we have argued previously~\cite{emn}, the appearance of
a term (\ref{generalemn})
is necessarily accompanied by
entropy production (decoherence) in the effective light-particle
system, at a rate given in general by (\ref{genentropy})~\cite{emn}.

\pr
In the $D$-brane case at hand, we specialize to the
subspace spanned by the tachyon modes 
(lowest-lying closed-string states), so that
(\ref{genentropy}) reduces to
\be
{dS^{light} \over dt} =  -\beta^T G_{TT} \beta ^T
\label{genentropy2}
\ee
where $T$ denotes a tachyon coupling.
In this application, the dominant contribution to
the conformal anomaly for the tachyon $\beta$ functions
comes from the interaction of the tachyon with the Goldstone 
modes $D_\epsilon,C_\epsilon$. We work in a renormalization 
scheme~\cite{zam} where
\be
   G_{TT} \sim 1 + O(T^2)
\label{metriczam}
\ee 
We now use the general expression 
$\beta ^i \sim C^i_{jk} g^j g^k + \dots$
for the $\beta$ function, where the $\dots$ denote higher-order
terms and the $C^i_{jk}=G^{im}C_{mjk}$, where the $C_{mjk}$ are
completely symmetric trilinear O.P.E. coefficients, i.e.,
three-point correlation functions $<V_iV_jV_k>$. In our $D$-brane case
the contributions of recoil to the conformal anomaly are of the form 
\be
 \beta ^T \simeq C^T_{C_gC_g} (u_i)^2 + \dots 
\label{confan}
\ee
to leading order in the matter deformations,
where $C_g$ denotes a bulk operator that represents the
virtual excitations appearing in $D$-brane recoil as discussed
in sections 2 and 3, having as couplings the 
recoil velocities $u_i$,
and the $C^T_{C_gC_g}$ denote appropriate O.P..E. coefficients~\footnote{An
explicit expression for $C_g$ is given later in (\ref{cgop}).}.
In this weak-tachyon perturbative treatment, the dominant contributions 
to the change in entropy (\ref{genentropy2})
of the light-particle subsystem is of order 
\be
dS^{light}/dt \propto \sum _T (u_i u^i)^2 (C^T_{C_gC_g})^2    
\qquad t \rightarrow \infty
\label{lightentr}
\ee
Below we shall 
express the bulk amplitude $C^T_{C_gC_g}$ in (\ref{lightentr})
in terms of open-to-closed scattering amplitudes 
of the open-string sector. This enables us to show that 
(\ref{lightentr}) is, to this order in the matter deformations, 
identical to (\ref{zamol}) but {\it opposite} in sign, 
thereby providing an explicit consistency check 
of the overall conformal invariance 
of the matter-plus-quantum-$D$-brane recoil system. 

\pr
As a step towards this demonstration, we first 
develop the proper definition of the light-state 
$\sigma$ model, so as to represent correctly the motion of 
low-lying closed-string modes in an `environment' of 
quantum-fluctuating $D$ branes. It is technically convenient 
for this purpose
first to re-express the $\sigma$-model for the $0$ brane itself
in the Liouville-dressed formalism,
using a Galilean transformation of the collective coordinates
of the $0$ brane. This, as we shall see, facilitates the analysis, 
as it provides a simple expression for $C_g$ in terms of a compact
representation of the $D$-brane boundary excitations (the $C$,$D$ pair).  
We rewrite the full recoil operator (\ref{recoil}) as
\be 
 V_{rec}= \epsilon   \int _{\partial \Sigma} d\tau \xi _i \partial _n \xi ^i 
\Theta _\epsilon (X^0) \qquad \xi _i \equiv y_i + \frac{u_i}{\epsilon}
X^0 
\label{xiform}
\ee
where the $\xi^i $ {\it do not} satisfy fixed Dirichlet boudnary
conditions. Using the analysis of quantum-mechanical uncertainties
of the $0$-brane subsystem~\cite{kogwheat}, which are
briefly discussed in the Appendix, c.f., eq. (\ref{qmu}) in particular,
we see that $\xi _i - y_i $ in (\ref{xiform}) expresses 
quantum uncertainty in the initial location of the 
$D$ brane, since $u_i/\epsilon $ is the minimum uncertainty 
obtained as a result of the action 
of the $D_\epsilon$ quantum-recoil operator.
This interpretation of (\ref{xiform}) may also be expressed
in the language of renormalization. We see
from eq. (\ref{velocity}) that
the effects of $\epsilon$ may be absorbed in an 
appropriate renormalization of the string coupling constant $g_s$,
and a corresponding renormalized recoil velocity 
$u_i^R \equiv u_i/\epsilon$. Eq. (\ref{qmu})
of the Appendix tells us that these renormalization effects result from  
quantum fluctuations in the location of the $0$-brane
collective coordinates. 
In this picture, which from 
a physical point of view corresponds to a Galilean transformation 
of the collective (spatial) coordinates
in a co-moving frame of velocity $u_i^R$,  
consistency with the renormalization-group approach requires $u_i^R$ 
to be exactly marginal, whilst $u_i$ will now have an anomalous 
dimension $du_i/dt=-(1/2t)u_i$, $t \sim 1/\epsilon ^2$, 
in agreement with 
the fact that the anomalous scaling dimension of the 
$C$ and $D$ operators 
is $-\epsilon ^2/2 \sim -1/(2t)$: $t \sim Log|L/a|^2$.  
In what follows, 
unless explicitly stated, we will suppress the superscript
when denoting the renormalized recoil velocity: $u_i^R \rightarrow u_i$.
In this interpretation, the $\epsilon \xi _i$ factor 
in (\ref{xiform}) may be considered as the minimum bound on the 
(quantum) uncertainties in the 
location of the $0$-brane in a co-moving (Galilean-transformed) frame.
According to the 
analysis in the Appendix (c.f., eq. (\ref{uncertc})), the uncertainty  
represented by the world-sheet zero mode of $\epsilon \xi _i$ 
is due to the action of the $C_\epsilon$ quantum-recoil 
operator, and is thereby related to the {\it 
stringy modes} of the $0$-brane~\footnote{Note that 
within our interpetation 
of $\epsilon ^{-2}$ as the target time $t$, this uncertainty
become time dependent. This is important for the correct
definition of the light-particle 
subsystem in this framework, as we discuss later on.}. 

\pr
Passing to the bulk formalism as in (\ref{bulkliouv}), 
expressing the coordinate $X^i$ in terms of $\xi ^i$, and 
identifying $\phi $ with $X^0$, one finds the following 
Liouville-dressed $\sigma$-model action:
\bea
 &~& S_\sigma^D = \int _\Sigma 
d^2z \frac{1}{4\pi \alpha '}[ 
-(1 - (u_i^R)^2)\partial _\alpha X^0 \partial ^\alpha X^0 
+ \partial _\alpha \xi ^i \partial ^\alpha \xi ^j \delta _{ij} 
+ \nn \\
&~&(-u_R^i + 
\epsilon ^2 \xi _i \Theta(X^0) )
\partial _\alpha X^0 \partial ^\alpha \xi ^i ]
+ \int _{\partial \Sigma} d\tau \epsilon \xi _i \partial _n \xi ^i 
\Theta (X^0) 
\label{liouvilledressing}
\eea
where the $\xi _i, X^0$ are independent $\sigma$-model fields, and 
$\Theta (X^0)$ is the Liouville-dressed $\Theta$ operator 
(\ref{thetaliouv}), which is conformal~\footnote{One may
renormalize the $X^0$ field in (\ref{liouvilledressing}) 
so that its kinetic term assumes the canonical 
form, in which case the off-diagonal metric term 
is multiplied by a factor $1/(1-(u_i^R)^2)$, which is 
almost unity for slow-moving $0$ branes, or
for recoiling branes in a low-energy scattering experiment.}.   
Notice that in the above approach $\epsilon$ is viewed as a 
parameter that is independent of the zero mode 
of the Liouville field $X^0$. The identification 
of $1/\epsilon ^2$ with $X^0$ should be made only 
at the very end of the computations. 
The off-diagonal metric in (\ref{liouvilledressing}) satisfies the 
generalized Weyl-invariance conditions for a transformation that
is singular at $X^0=0$. Moreover, 
the curvature scalar constructed out of the coordinates $\xi_i, X^0$,
exhibits the temporal singularity (\ref{scalrcurv}) at $X^0=0$, 
$R \ni -\epsilon ^4 \xi _i \xi ^i \delta (X^0) \Theta (X^0)$
due to the effects of $D$-brane recoil, which induces the 
above-mentioned uncertainty $\epsilon \xi $ in the location of the brane.  

\pr
In the formalism (\ref{liouvilledressing}),
the recoiling boundary quantum state is characterized by
a single Goldstone-mode operator $\cal C$ of the $C$ type:
\be
S^D_{\sigma} \ni \int _{\partial \Sigma } d\tau {\cal C} (\tau)
\equiv \int _{\partial \Sigma} d\tau \epsilon \xi _i \partial _n \xi ^i 
\Theta (X^0) 
\label{cope}
\ee
which incorporates 
the $C$,$D$ operator mixing discussed in section 2,
by virtue of the transformed $\xi$ variable.
On the other hand, a generic macroscopic $D$-brane boundary state $|a>$ 
is described by an operator of the form 
(\ref{cope}) without the $\Theta (X^0)$ term 
and with $\epsilon~\xi$ replaced 
by a macroscopic (fixed) coordinate $y_i$. This is 
not exhibited explicitly in (\ref{liouvilledressing}), 
but it should be understood in the following.
We recall that in the modern approach to $D$ branes~\cite{dbranes}, 
the Dirichlet boundary conditions for the collective coordinates
are viewed as reflecting appropriate gauge-field background excitations 
of ordinary open-string $\sigma$ models, 
with all the coordinates of the string
obeying standard Neumann (closed-string) boundary 
conditions~\cite{dbranes,bachas}. 
This will always be implicit in our subsequent manipulations 
based on the action (\ref{liouvilledressing}). 
The $\sigma$-model action $S_\sigma ^D$ (\ref{liouvilledressing})
describes the $0$ brane, viewed as a subsystem for which the 
scattered matter has been integrated out. 
Hence the Liouville field/time $X^0$ appearing in (\ref{liouvilledressing})
represents collectively the matter effects which induce violations 
of conformal invariance that are restored 
by the gravitational (Liouville) dressing.

\pr
We now use the above formalism (\ref{liouvilledressing}) to 
construct the matter-subsystem $\sigma$ model,
which enables us to evaluate the right-hand side 
of (\ref{genentropy2}). In the bulk formalism, our 
$\sigma$ model describing the 
matter subsystem is {\it defined} by 
(\ref{liouvilledressing}), upon {\it dropping} the boundary term
(\ref{cope}) which encodes the environment  
of quantum-recoil fluctuations of the $D$ brane,  
and {\it adding} the closed-string tachyon deformation: 
\be 
       T(\xi, X^0) = \int d^Dk d\omega e^{ik_i \xi ^i } e^{i\omega X^0}
\label{closedtach}
\ee
where $D$ is the target-space dimension.
This definition makes physical sense for the following reason:
As we mentioned previously, 
in the closed-string formalism one may view
a $D$ brane located originally at $y_i$ 
as being created by an appropriate operator
which creates the macroscopic boundary state of the brane $|a>$. 
On the other hand, a {\it low-energy} observer is obliged to 
`average out' the time-dependent {\it stringy} effects due to
quantum uncertainties in the location of the brane associated with
the virtual recoil excitations, that are expressed via
the boundary term (\ref{cope}). In $\sigma$-model language, 
such an averaging-out procedure is equivalent to ignoring 
the corresponding deformations of the world-sheet action.   
In this framework, the bulk operator $C_g$ appearing in (\ref{confan}), 
which describes the quantum
$D$-brane recoil excitations from a matter-theory 
point of view, and hence is retained in the 
action for the light-matter subsystem,  
may be identified with the following deformation: 
\be
S^D_{\sigma} \ni \int _\Sigma d^2 z C_g (z,{\bar z})  
\equiv \epsilon ^2 \int _\Sigma d^2z [\xi _i \Theta(X^0) 
\partial _\alpha X^0 \partial ^\alpha \xi ^i ] 
\label{cgop}
\ee
This completes our specification of the matter $\sigma$ model.

\pr
We now check the mathematical consistency 
of this definition for the matter subsystem
by reproducing the entropy change (\ref{zamol})
through (\ref{genentropy2}). 
First we show that the operator $C_g$ (\ref{cgop}) contains the bulk 
form of the recoil $D$ operator discussed in section 2,
by splitting the field $\xi ^i $ into zero-mode and fluctuation
parts:
\be 
\xi _i = \xi _{0,i} + {\hat \xi }_i (z,{\bar z}) 
\label{xifluct}
\ee
where $\xi _{0,i} $ is independent of the world-sheet coordinates. 
The operator $C_g$ is now seen to be equivalent to 
\be 
\int C_g \simeq \epsilon^2 \xi _0 
\int _{\Sigma} d^2 \partial _\alpha (\int \frac{dq}{iq^2} e^{iq~X^0})
\partial ^\alpha \xi^i  + \dots 
\label{cgdop}
\ee
where $\dots $ denote the ${\hat \xi }$-dependent parts. 
Clearly, if the world sheet has a boundary, 
the operator exhibited in (\ref{cgdop}) 
becomes the $D$-brane recoil operator 
$\epsilon ^2 \xi _0 
\int _{\partial \Sigma} d\tau D (\tau) + \dots $, 
where the $\dots $ denote world-sheet terms that vanish on shell. 
In this spirit, the `coupling' $\epsilon ^2 \xi _0 $ in (\ref{cgdop}),
which has the dimensions of velocity, 
plays the r\^ole  of the renormalized recoil velocity $u_i$ 
of the brane. These $\xi _0$ zero-mode parts of 
$C_g$ contribute the dominant divergent terms in the conformal 
anomaly $(\ref{confan})$, and we now concentrate on them.

\pr
There is a formal connection between the closed (bulk) 
and open (boundary) string theories~\cite{cardy}, 
which leads
to a determination of (\ref{lightentr}) in terms of quantities in the 
open-string formalism, which provides the advertized connection 
with (\ref{zamol}). We recall that
the matrix element $C_{C_gC_g}^T$ describes - to 
leading order in the matter deformation - the interaction of the 
bulk gravitational modes of the $D$ brane, expressing 
quantum recoil,  
with the light matter in the closed-string channel~\cite{periwal,emnd}. 
According to ref. \cite{cardy}, such bulk amplitudes may be 
expressed as `squares' of open-to-closed string amplitudes 
containing excitations on the (conformal) boundary state $|a>$. 
The pertinent relation 
is of the form~\footnote{For a rigorous proof of (\ref{cg}) 
in the presence of logarithmic operators, see~\cite{kogwheat2}.}:
\be
    \sum _{T} C^T _{C_gC_g} \propto (C^{a}_{C_g {\cal C}})^2 
\label{cg}
\ee
with $C_g$ given by (\ref{cgop}) and $\cal C$ by (\ref{cope}).

\pr
We now notice that the sum over tachyon modes 
in (\ref{cg}) actually reduces to a single term in  the string case,
as a result of the path integration 
of the zero-mode of $X^0$. This can be seen easily by  
using (\ref{closedtach}) 
for the tachyon mode 
and taking into account the specific form of the dominant
$\xi _0$ zero-mode 
part of the $C_g$ operators (\ref{cgdop}):
$C_q \propto \int dq \frac{1}{q^2} e^{iqX^0} $. 
It is then straightforward to see that the integration over the zero mode
of $X^0$ simply imposes energy conservation:
$\omega + q + q'=0$
for the three-point amplitudes in (\ref{cg}), thereby reducing
the sum over tachyon modes $T$ (i.e., over $\omega$) to a single term 
which is uniquely specified in terms of the $C_g$. 
This allows (\ref{cg}) to be inserted in the right-hand side of 
(\ref{lightentr}).

\pr
We now recall the fact, mentioned above, that 
the dominant contributions to (\ref{confan}) 
arise from the zero-mode parts of $C_g$.
Equivalently, in the open-string formalism, the dominant
contributions appear when the operator $C_g$ is near the boundary,
and its one-point function diverges~\cite{periwal,kogwheat,emnd}.
Using (\ref{cgdop}), we then see easily that 
the amplitude $C^{a}_{C_g {\cal C}}$ is essentially   
$<CCD>$. Since the one-point function of the 
$D$ operator diverges
like $1/\epsilon$~\cite{periwal,kogwheat} 
near the boundary (\ref{divoper}),
and the two-point O.P.E. of $CC$ vanishes as~\cite{kogwheat}
$\epsilon ^2$ (\ref{cc3}),
it follows that the leading short-distance singularity
of the above-mentioned 
three-point function, when the arguments of the three 
operators are all behaving ${\cal O}(\epsilon)$, is:
\be 
C_{C_g {\cal C}}^a \sim \epsilon 
\label{copeps}
\ee  
which implies that 
\be 
C^T_{C_gC_g} \sim (C^a_{C_g {\cal C}})^2 \sim \epsilon ^2 
\label{ctgg}
\ee
Identifying  $1/\epsilon ^2 \sim t$, where  $t$ 
is the target time/renormalization-group scale,  
we obtain the following result for the rate of change 
(\ref{lightentr}) of the entropy of the light subsystem:
\be
dS^{light}/dt \propto -(u_i u^{i} )^2 \frac{1}{t^2} \qquad t \rightarrow \infty    
\label{fentr}
\ee
which is {\it identical} to 
the result (\ref{zamol}), but {\it opposite} in sign.
Thus, for the light-particle system,
the direction of time $t$ must be taken {\it opposite} to 
that of the Liouville renormalization-group flow, in order 
to have increasing entropy. This is consistent with the  
fact that the total entropy of the brane-plus-matter system 
remains constant as a consequence of conformal invariance. A similar 
situation has been argued to hold in the non-critical 
Liouville analysis of the string
black hole. The above analysis 
completes the consistency check on our definition 
of the light-state subsystem, to lowest non-trivial 
order in matter deformations. 

\pr
The expression (\ref{fentr}) is finite and
independent of the area of the 
$D$-brane collective-coordinate surface. Hence, the Bekenstein-Hawking
area law is not valid for the information-theoretic entropy associated
with the quantum fluctuations of a recoiling $D$ brane. We recall that
the area law is semiclassical, being associated with a tree-level
treatment of the gravitational background, which is described in our
case by a $D$-brane configuration. The semiclassical treatment ignores
recoil, and simply expresses the entropy of the subsystem of tachyons 
obtained by integrating over the tachyon degrees of freedom behind 
the collective-coordinate surface of a classical $D$-brane configuration.
This is a geometric term which is also present 
in standard flat-space local field theories~\cite{srednicki}.
Our analysis goes beyond this by considering quantum fluctuations
in the background induced by recoil during a scattering process.
The situation is analogous to that studied in~\cite{emnw},
where the information-theoretic content of  
quantum fluctuations in a scalar field at the horizon of a
four-dimensional black hole in General Relativity
also violated the area law. 
Recent claims~\cite{susskind} 
to have derived the Bekenstein-Hawking  
law for the string black-hole entropy, 
considering the horizon of a stringy black hole as a $D$ brane,
apply only at the semiclassical level.

\pr
The order of magnitude of the decoherence  
term (\ref{fentr}) is the same as that of the 
non-Hamiltonian term $\nd{\delta H}$ in the 
evolution equation of the density matrix of the 
light subsystem. This is due to the fact that, for asymptotic 
$t \rightarrow \infty$ where the non-critical string
is near its equilibrium (critical) point, the commutator 
$i[g^i, \rho] \sim i[g^i, e^{-H}] =O[ \partial _t g^i =\beta ^i ]$, as 
a result of the canonical formalism for the couplings
$g^i$~\cite{emninfl,emnd}.  
According to the discussion above, the decoherence 
of the light-state subsystem is determined by the
recoil velocity of the collective-coordinate surface of the 
$D$ brane induced during scattering. This is given by 
(\ref{velocity}), and
can be at most of the same order 
of magnitude as $g_sE$, where $E$ is a typical energy of the 
propagating closed-string state, which
plays the r\^ole of the low-energy propagating particle
treated in~\cite{emn}). Equation (\ref{fentr}) tells us that,
for long times $t \rightarrow \infty$, the dominant 
decoherence effects in the evolution equation 
of the density matrix are  
of order $\frac{g_s^4}{M_s^4}E^4$, where $M_s$ is the string scale. 
To make this estimate, we take into account the fact that
the evolution parameter on the world sheet scale is
measured in string units $\sqrt{\alpha '}$, where $\alpha '
=1/M_s^2$ is the string Regge slope. 

\pr

\section{Conclusions}

\pr
We have argued in this paper that $D$ branes provide another
example of a model of quantum gravity in which the effective
theory of the propagating light-particle states obeys a
modified quantum Liouville equation of the type (\ref{ehns}),
following the previous examples of $1+1$-dimensional string
black holes~\cite{emn} and a scalar
field outside a four-dimensional black hole with Yang-Mills interactions.
As in these previous cases, time is interpreted as a renormalization-group
scale parameter, which may be identified in the $D$-brane and string
black-hole cases with the zero mode of the Liouville field on the
world sheet. In these two string models, there is no question of
modifying quantum mechanics on the world sheet. In both cases, the
light degrees of freedom are to be regarded as truncated open systems,
and the modification (\ref{ehns}) of the quantum Liouville equation
can be traced to the need to treat the interaction with the
black-hole environment as a source term in the world-sheet equation
of motion for the light-particle system, pushing it away from
criticality~\cite{emnd}.

\pr
We take this opportunity to clarify where we differ from views
espoused elsewhere~\cite{polchinskireview,banks} on the
possible maintenance of quantum coherence. In our view, the fact that
the conventional Hawking-Bekenstein black-hole entropy can be identified
with the number of distinguish$able$ $D$-brane states does not mean
that these states are necessarily distinguish$ed$ in a realistic
experiment: for example, the infinite set of Aharonov-Bohm phase
measurements described in~\cite{emnmeas,banks} are $not$ feasible
in practice. This is why one must sum over the $un$observed degrees
of freedom, which are represented in the present case by the
$quantum$ excitations of the $D$ brane, that cause a deviation from the
Hawking-Bekenstein area law and were not studied in~\cite{banks}. It
is convenient technically to perform the
sum over unseen states using the Liouville field, whose
zero mode we interpret as time, as confirmed in this
$D$-brane analysis.

\pr
During the course of this analysis, we have indicated various steps
where a more rigorous treatment is desirable and in
preparation~\cite{kogwheat2}. There are also some important
physical issues that should be addressed in future work on this
$D$-brane model. One is the possible r\^ole of infinite-dimensional
symmetries such as the $W_{\infty}$ discussed in connection with
the $1+1$-dimensional black-hole model. We have argued 
previously~\cite{emnwhair,emn}
that the infinite-dimensional Cartan subalgebra of $W_{\infty}$
provides an infinite set of conserved charges that are available
to label black-hole states with $W$ hair. We have further argued
that a finite low-energy experiment is unable to measure
all the $W$ hair~\cite{emn}, and that the corresponding inevitable
leakage of $W$ quantum numbers between the light-particle subsystem and
the black-hole background is a measure of the rate of information loss
inherent in (\ref{ehns}). Previously, we have presented formal
evidence for this leakage in the $1+1$-dimensional black-hole
model~\cite{emn}, 
and exhibited analogues of these $W$ charges in a
higher-dimensional $D$-brane model. The possible r\^ole of such
charges in the model studied here deserves further investigation.
We are optimistic that such $W$ charges do indeed exist, since
it is known~\cite{shaf} that a world-sheet 
$W_{\infty}$ algebra~\cite{pope} 
is present
in any model with logarithmic operators $C,D$, such as this one.
Moreover, the singularity (\ref{scalrcurv})
at $X^0 = 0$ in this model is
effectively $1+1$ dimensional, leading us to expect that it may
have an underlying algebraic structure similar to the $1+1$-dimensional
black-hole example. However, these points require further analysis.

\pr
Another area worthy of deeper analysis is the creation and
annihilation of $D$ branes. In this paper, we have 
a pre-existing $D$-brane background, and discussed the
formalism for quantum excitations on its surface. However, we
expect that much of the machinery developed here could also be
used to discuss the production, annihilation and decay of $D$
branes. In particular, the $\Theta (X^0)$ and singularity structure 
of the recoil operator (\ref{recoil}) is well adapted 
to the creation of a $D$-brane pair. However, an extension in necessary 
to describe 
``$D$-brane foam", i.e., the treatment of virtual $D$-brane
fluctuations in the space-time background~\footnote{This was achieved 
in the $(1+1)$-dimensional-black-hole case using a world-sheet 
valley approach~\cite{emnval}.}. 
We have developed
previously~\cite{emnval} the corresponding description of $1+1$-dimensional
black holes in terms of monopoles on the world sheet. Extending
the present analysis to virtual $D$ branes is, however, a non-trivial
technical problem.

\pr
In the absence of an appropriate treatment, we can nevertheless
offer some intriguing speculations. Let us suppose, as was found
in this paper, that the information lost in any encounter with a
microscopic $D$ brane is $\propto E^4 M_P^{-4}$, where $E$ is a typical 
energy scale of the light state. Let us further
hypothesize that the density of virtual $D$-brane excitations of
the space-time background is ${\cal O}(M_P^3)$~\footnote{We do not
distinguish for this heuristic argument between $M_P, 
M_s = 1/\sqrt{\alpha'}$,
or any other gravitational scale in string theory.}. On the other hand, 
the 
low-energy particle cross section $\sigma$ 
is assumed to be proportional to 
$1/E^2$~\cite{emohn}. Thus, 
if the
speculations in the two preceding sentences were correct, we would
infer that $\nd{\delta H} \propto 1/M_P$, where by dimensional analysis
the numerator would be ${\cal O}(E^2)$, where $E$ is a characteristic
energy scale of the light-particle system. This heuristic argument
is consistent in order of magnitude with estimates made previously
in the contexts of the $1+1$-dimensional black hole and the
four-dimensional quantum gravity model. Remarkably, it is also
of the same order as the sensitivity of the neutral kaon~\cite{emnkaon}
system to any
such possible deviations from the canonical quantum Liouville
equation. Therefore we cannot yet exclude the possibility that the
type of decoherence effect discussed in this paper might have some
phenomenological relevance, however far-fetched this may seem.

\newpage
\pr
{\Large {\bf Acknowledgements}}
\pr
It is a pleasure to acknowledge useful discussions 
with G. Amelino-Camelia, I. Kogan, F. Lizzi and 
J.F. Wheater. N.E.M. wishes to thank the CERN Theory 
Division for its hospitality during the final 
stages of this work. The work of D.V.N. 
has been supported in part by DOE grant 
DE-FG05-91-ER-40633. 
\pr

{\bf Appendix: $D$-brane Decoherence and Position Uncertainty}

\pr
Additional support for our picture of information 
loss and decoherence induced by the recoil of $D$ branes
is provided by a closer look at the modified uncertainty 
principle for $D$ particles which was discussed previously 
in ref.~\cite{kogwheat}.
As was discussed there, the r\^ole of the logarithmic operators
describing recoil is crucial in a derivation of an 
uncertainty principle for $D$ branes. The $D$ operator
provides the conventional quantum-mechanical part of the 
uncertainty, whilst the $C$ operator 
is associated with intrinsically stringy parts. 
To be more precise, if one
considers the scattering of light closed-string state,
playing the r\^ole of a `detector', off a $D$ brane, 
and assumes that the momentum of the detector is known precisely before 
and after the scattering, one arrives at the conclusion that 
there is an uncertainty in the energy of the $D$ brane which 
is of order $\hbar \epsilon$. 

\pr
This conclusion was reached~\cite{kogwheat} 
by first observing that the regulated $\Theta _\epsilon (X^0)$
operator appearing in the definition of the recoil operators
(\ref{recoil}) 
may be written as $\Theta _\epsilon \sim \Theta (X^0)e^{\epsilon
X^0}$; from this it follows that the recoil (measurement) 
takes a finite time $\Delta \tau \sim 1/\epsilon $, thereby implying 
an uncertainty in the total energy of order $\hbar \epsilon$.   
The kinematical analysis of ref. \cite{kogwheat}
established the following relation: 
\be 
 \Delta P^i \sim \hbar \epsilon/u^i 
\label{uncert}
\ee
where $\Delta P^i$ denotes the uncertainty in the 
$D$-brane momentum. 
Given that the collective coordinate of the $D$ brane 
starts growing as $u_i t$ (\ref{scale2}), the uncertainty  
(\ref{uncert}) implies an uncertainty due to the $D_\epsilon $ 
recoil operator in the position of the collective coordinate 
of the $D$ brane of order 
\be
  \Delta _{D_\epsilon} X^i \sim u_i \Delta t =
u_i \frac{1}{\epsilon} = \frac{\hbar}{\Delta P^i} 
\label{qmu}
\ee
which is a quantum-mechanical point-like particle 
uncertainty betweeen coordinates and momenta. 

\pr
The essentially stringy parts of the uncertainty 
are obtained from the $C_\epsilon $ operator, which,
due to its specific form (\ref{cop}),
yields a lower bound on the uncertainty 
of the coordinate $y^i$ of order $y^i \epsilon $.
Using (\ref{uncert}) one obtains 
\be 
 \Delta _{C_\epsilon} X^i \sim \frac{y^i u_i}{\hbar} \Delta P^i 
\label{uncertc}
\ee
Thus, the total uncertainty in the collective spatial coordinate 
of a recoiling $D$ brane, due to the combined action of the 
$C$ and $D$ logarithmic pair , is 
\be 
  \Delta _{total}  X^i \sim \frac{\hbar}{ \Delta P^i} 
+ \frac{y^i u_i}{\hbar} \Delta P^i 
\label{deltap}
\ee
It is important to stress once again that, due to the mixing between 
the operators $C$ and $D$, one cannot evade the influence of the 
$C$ operator, and hence a `stringy' 
form of the uncertainty principle for a $D$ brane~\cite{kogwheat}. 

\pr
For our purposes, it is important to notice the dependence 
of the second term in (\ref{deltap}) on the coordinate $y^i$.
Minimizing the right-hand-side of the uncertainty (\ref{deltap})  
with respect to $\Delta P^i$, one obtains a lower bound on $\Delta X^i$ 
of the form:
\be 
   (\Delta X^i)_{min} \sim 2 \sqrt{u_iy^i}=4\sqrt{2\sqrt{2}
\pi g_s~(k_1 + k_2)_i~y^i}
\label{totalunc}
\ee
The $\sqrt{y^i}$ dependence is reminiscent of the 
measurability bounds occuring in the non-critical 
Liouville string approach to 
target time~\cite{amelino}. As discussed there, 
the result (\ref{totalunc}) is compatible with the decoherence  
(\ref{zamol}) implied by the effects
of the recoil operator $C_\epsilon$. 
This similarity should not have come as a surprise, given the discussion 
above on the marginal perturbation of the recoiling $D$ brane from 
a conformal point, and the generic connection of soliton 
backgrounds in string theory with Liouville strings~\cite{emnd}.



\begin{figure}
\hglue2.0cm
\vglue-1.7cm
\epsfig{figure=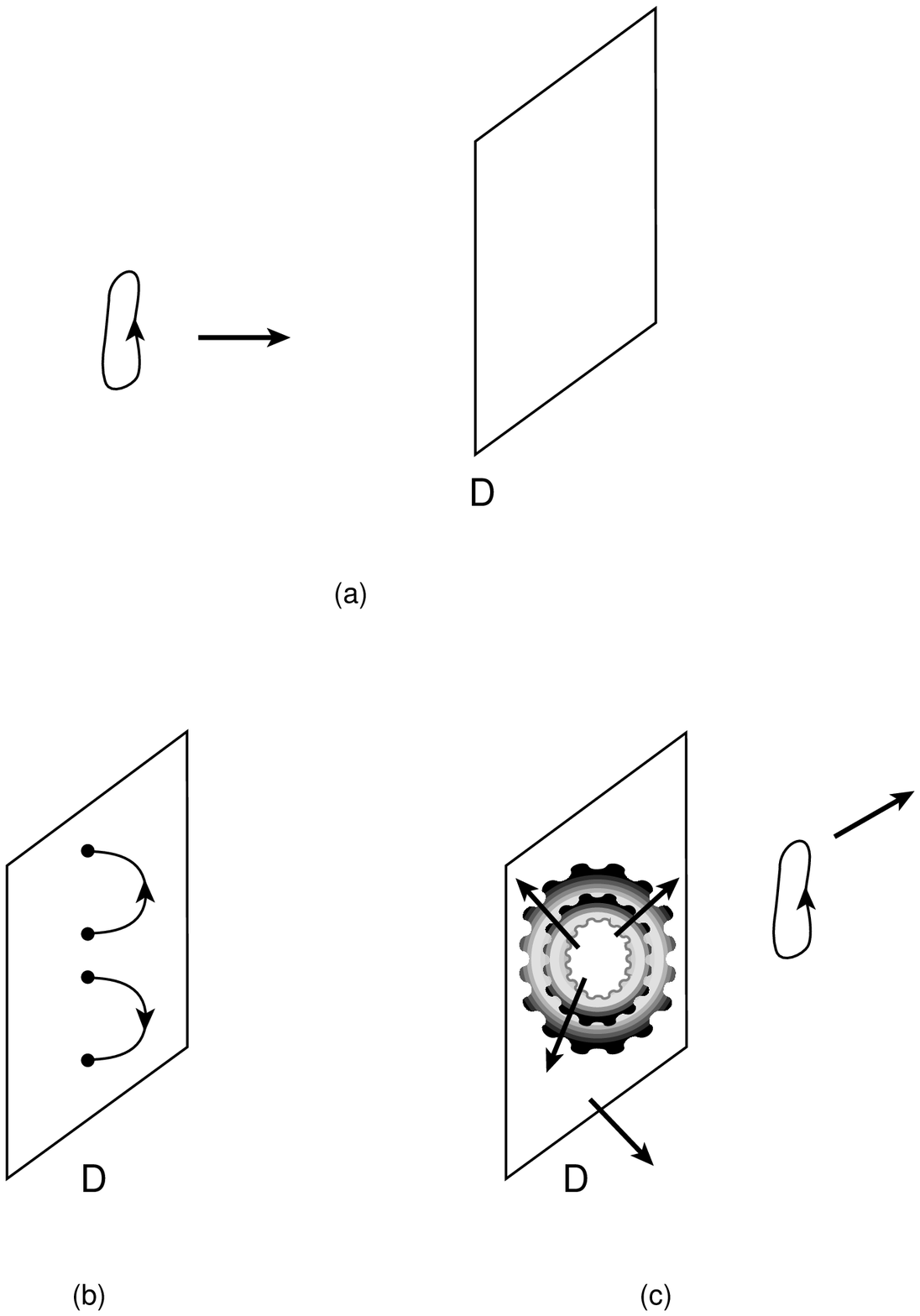,width=14.cm}
\vspace{0.2cm}
\caption[]{ Scattering of a low-energy
closed-string state off a $D$ brane: (a) asymptotic past, 
(b) time of impact ($X^0=0$) with 
trapping of the string state on the $D$-brane surface by 
a split into two open-string excitations, and 
(c) asymptotic future, after two open strings recombine to emit a 
closed-string state, while the 
$D$ brane recoils with finite velocity and its internal 
state fluctuates}
\end{figure}

\pr
\pr
\begin{figure}
\hglue3.5cm
\epsfig{figure=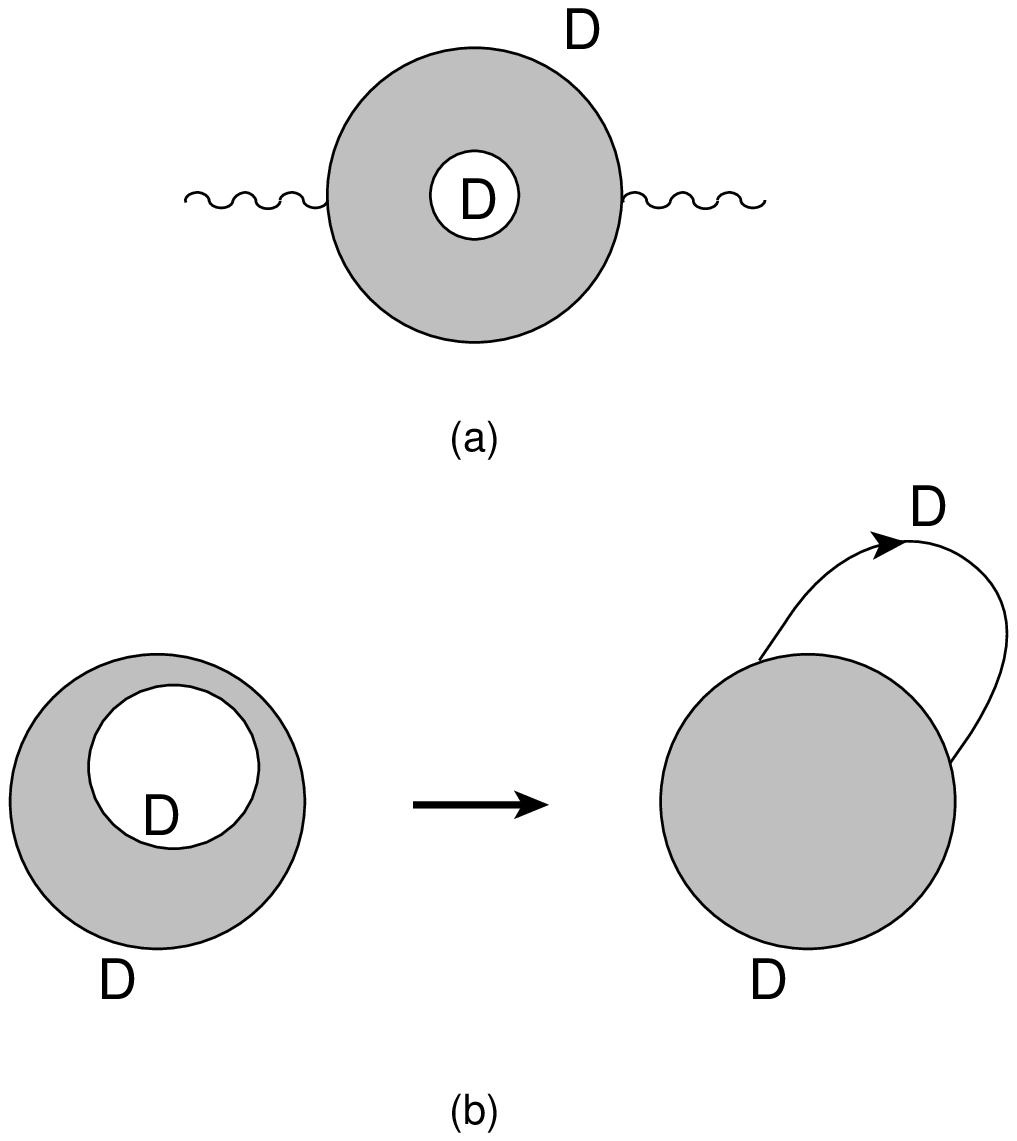,width=11cm}
\caption[] { (a) : World-sheet
annulus diagram 
for the leading quantum correction to the 
propagation of a string state in a $D$-brane background, and 
(b) the pinched annulus configuration which is the dominant divergent 
contribution to the quantum recoil.}
\end{figure} 
\end{document}